\documentclass[amsmath,amssymb,twocolumn,aps,prb,superscriptaddress,dvipdfmx]{revtex4-2}

\usepackage{graphicx}
\usepackage{bm}
\usepackage{braket}
\usepackage{color}

\usepackage{amsmath}
\usepackage{mathtools}
\usepackage{empheq}
\usepackage{ulem}

\usepackage[colorlinks=true,linkcolor=blue,linkcolor=blue,urlcolor=blue,citecolor=blue]{hyperref}

\newcommand{\daverage}[1]{\ensuremath{\langle \langle #1\rangle \rangle} }  
\newcommand{\average}[1]{\ensuremath{\langle#1\rangle} }  
\newcommand{\Equref}[1]{Eq.~(\ref{#1})}  
\newcommand{\Figref}[1]{Fig.~\ref{#1}}  

\newcommand{\bmH}{{\bm H}}

\newcommand{\bmm}{{\bm m}}
\newcommand{\bmn}{{\bm n}}

\newcommand{\bms}{{\bm s}}

\newcommand{\kB}{k_{\rm B}}

\begin{document}

\title{Antiferromagnetic spin Seebeck effect across the spin-flop transition:\\
A stochastic Ginzburg-Landau simulation}

\author{Yutaka Yamamoto}
\affiliation{Department of Physics, Okayama University, Okayama 700-8530, Japan}
\author{Masanori Ichioka}
\affiliation{Research Institute for Interdisciplinary Science, Okayama University, Okayama 700-8530, Japan}
\affiliation{Department of Physics, Okayama University, Okayama 700-8530, Japan}
\author{Hiroto Adachi}
\affiliation{Research Institute for Interdisciplinary Science, Okayama University, Okayama 700-8530, Japan}
\affiliation{Department of Physics, Okayama University, Okayama 700-8530, Japan}

\date{\today}

\begin{abstract}

  We investigate the antiferromagnetic spin Seebeck effect across the spin-flop transition in a numerical simulation based on the time-dependent Ginzburg-Landau equation for a bilayer of a uniaxial insulating antiferromagnet and an adjacent metal. By directly simulating the rate of change of the conduction-electron spin density $\bms$ in the adjacent metal layer, we demonstrate that a sign reversal of the antiferromagnetic spin Seebeck effect across the spin-flop transition occurs when the interfacial coupling of $\bms$ to the staggered magnetization $\bmn$ of the antiferromagnet dominates, whereas no sign reversal appears when the interfacial coupling of $\bms$ to the magnetization $\bmm$ dominates. Moreover, we show that the sign reversal is influenced by the degree of spin dephasing in the metal layer. Our result indicates that the sign reversal is not a generic property of a simple uniaxial antiferromagnet, but controlled by microscopic details of the exchange coupling at the interface and the spin dephasing in the metal layer.

\end{abstract}

\maketitle
%
%
\section{\label{sec:level1}INTRODUCTION}
Although an antiferromagnet violates the rotational symmetry in spin space~\cite{Anderson1963}, it does not show net magnetization under zero magnetic field~\cite{Nagamiya1955}. Because of this magnetically silent feature on a macroscopic scale, in comparison to ferromagnets, antiferromagnets had not been a major player in practical applications of magnets~\cite{Coey2009}. However, the advent of spintronic technology put a new twist on this practice because manipulation of spins on the sub-atomic scale allows us to harness the staggered moments in antiferromagnets. Indeed, the discovery of exchange bias effect has opened up industrial applications of antiferromagnets~\cite{Coey2009}. Since then, especially over the last 10 years, substantial progress has been made in antiferromagnetic (AF) spintronics~\cite{Jungwirth2016, Zelezny2018, Baltz2018}, as exemplified by an extraordinarily good spin transport property~\cite{Wang2014, Lin2016, Lebrun2018a, Yuan2018, Hahn2014} and an electrical manipulation of the N\'{e}el ordered parameter~\cite{Wadley2016, Bodnar2018}. Therefore, it is of significant importance to investigate the spin current physics in antiferromagnets.

 For the investigation of spin current physics, one choice is to use spin Seebeck effect (SSE) as it is a simple and versatile means for generating the spin current from a temperature bias~\cite{Adachi2013, Uchida2016}. Since discovered in a ferromagnet~\cite{Uchida2008}, the SSE has been extensively studied in a series of insulating ferrimagnets \cite{Uchida2010c,Uchida2016,Geprags2016}. The SSE has recently been extended to insulating antiferromagnets such as $\rm{Cr_2O_3}$~\cite{Seki2015}, $\rm{MnF_2}$~\cite{Wu2016}, $\rm{NiO}$~\cite{Holanda2017a,Hoogeboom2021}, $\alpha$-$\rm{Cu_2V_2O_7}$~\cite{Shiomi2017}, and $\rm{FeF_2}$~\cite{Li2019}. This AF SSE offers a nice playground to investigate the interplay of spin current and antiferromagnetism~\cite{Takei2015,Tatara2019, Shen2019, Troncoso2020}.

 Recently, the AF SSE was observed in $\rm{Cr_2O_3}$ with a simultaneous measurement of the AF spin pumping~\cite{Li2020}. This material is a uniaxial antiferromagnet, and under a magnetic field parallel to the anisotropy field the system exhibits a spin-flop (SF) transition, i.e., a spin structural transition from parallel to perpendicular configurations with respect to the AF easy axis~\cite{Coey2009}. A surprising feature of the result is that the sign of the SSE signal is reversed across the SF transition. Subsequently, a similar behavior is reported in $\alpha$-$\rm{Fe_2O_3}$~\cite{Yuan2020}.

 Experimentally, in previous AF SSE measurements such a sign reversal was {\it not} observed either in $\rm{Cr_2O_3}$~\cite{Seki2015} or in $\rm{MnF_2}$~\cite{Wu2016}. On theoretical side, the sign of the AF SSE below the SF transition was argued to be the same as the ferromagnetic SSE in Ref.~\cite{Hirobe2017} using the Holstein-Primakoff magnons valid at low temperatures (see supplemental material therein). The same conclusion was derived in Ref.~\cite{Yamamoto2019} using the Ginzburg-Landau model valid near the N\'{e}el temperature. By contrast, it was argued quite recently~\cite{Reitz2020} that the sign of the AF SSE below the SF transition is opposite to that of the ferromagnetic SSE. Given these conflicting results, clarification of whether or not the sign reversal of the SSE is a generic property of a simple uniaxial antiferromagnet, is an urgent issue.

 In this paper we present a numerical investigation of the AF SSE across the SF transition. We simulate the time-dependent Ginzburg-Landau (TDGL) equation for an insulating uniaxial antiferromagnet with thermal noise. In thermal equilibrium, the Ginzburg-Landau model is well known to describe the SF transition near the N\'{e}el temperature (see Sec.~48 of Ref.~\cite{Landau1984}). The TDGL equation which we employ here~\cite{Yamamoto2019, Freedman1976, Halperin1976} is an appropriate dynamical generalization of the equilibrium Ginzburg-Landau model. By simulating the spin current pumped into the adjacent paramagnetic (PM) metal that acts as a spin-Hall electrode, we examine temperature and magnetic field dependences of the AF SSE across the SF transition. Then we show that the sign of the AF SSE across the SF transition is determined by the microscopic details of the interfacial exchange coupling between the antiferromagnet and the spin-Hall electrode. Namely, when the conduction-electron spin density $\bms$ of the spin-Hall electrode dominantly couples to the staggered magnetization $\bmn$ of the antiferromagnet (we term this interaction ``N\'{e}el coupling'', which is allowed for interfaces as exemplified by the exchange bias effect), the AF SSE reverses its sign across the SF transition. By contrast, when $\bms$ dominantly couples to the magnetization $\bmm$ (we term this interaction ``magnetic coupling''), no sign reversal appears. Moreover, the appearance of the sign reversal also requires the condition that the spin dephasing of $\bms$ is not too strong (for a quantitative discussion, see Sec.~\ref{sec:level5}). Based on these results, we propose a possible mechanism for the sign reversal. That is, the presence of the sign reversal in recent experiments for $\rm{Cr_2O_3}$~\cite{Li2020} and $\alpha$-$\rm{Fe_2O_3}$~\cite{Yuan2020} may be understood as a result of the sizable N\'{e}el coupling at the interface, whereas its absence in previous experiments for $\rm{Cr_2O_3}$~\cite{Seki2015} and $\rm{MnF_2}$~\cite{Wu2016} may be due to the  dominant role of the magnetic coupling. 

 The rest of this paper is organized as follows. In Sec.~\ref{sec:level2} we describe our model, i.e., the TDGL for a uniaxial antiferromagnet that exhibits the SF transition. In Sec.~\ref{sec:level3} we calculate equilibrium phase diagram of the system and perform a numerical simulation of the antiferromagnetic resonance (AFMR), in order to check if our model can appropriately describe known physical properties of a uniaxial antiferromagnet. In Sec.~\ref{sec:level4}, we perform a numerical simulation of the AF SSE to investigate the sign reversal phenomenon across the SF transition. The results are discussed in Sec.~\ref{sec:level5}, and a conclusion is given in Sec.~\ref{sec:level6}.
 %
%
 \section{\label{sec:level2} MODEL}
\begin{figure}[t]
  \begin{center}
    \includegraphics[scale=0.4]{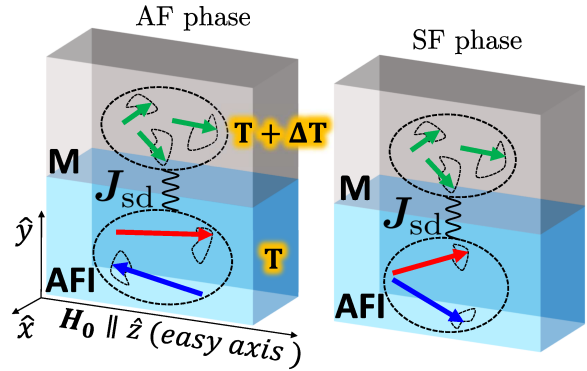} 
  \end{center}
\caption{\label{fig:bilayer} Schematic illustration considered in this paper for the AF SSE. AFI and M refer to an antiferromagnetic insulator and a metal, respectively, and the interfacial exchange interaction is denoted as $J_{\rm sd}$, which corresponds to either $J_\text{sd}=J_m$ (magnetic coupling) or $J_n$ (N\'{e}el coupling). }
\end{figure}

In this section, we describe the TDGL model used in this paper. We consider a bilayer composed of an antiferromagnetic insulator (AFI) with its temperature $T$ and a metal (M) playing the role of spin-Hall electrode with its temperature $T+\Delta T$, as shown in Fig.~\ref{fig:bilayer}.

First, we focus on the AFI layer. For AFI, we use the Ginzburg-Landau free energy of the following form \cite{Yamamoto2019, Landau1984}:
\begin{eqnarray}
  \label{eq:Free_Energy1}
  F_{\mathrm{AFI}} &=& \epsilon _0 v_0 \biggl\{ \frac{u_2}{2}\bm{n}^2 + \frac{u_4}{4}(\bm{n}^2)^2 + \frac{K_0}{2}(\bm{n}\times \bf{ \hat{z} } )^2  \nonumber\\
  && + \frac{K_1}{2}(\bm{n}\cdot {\bf \hat{x}})^2 + \frac{D}{2}(\bm{n}\cdot \bm{m})^2 +  \frac{D'}{2} \bm{m}^2\bm{n}^2 \nonumber \\
  && + \frac{r_0}{2} \bm{m}^2-\frac{\bm{H}_{\text{ex}}}{\mathfrak{h}_0}\cdot \bm{m} \biggl\},
\end{eqnarray}
where $\bm{m}$ and $\bm{n}$ are respectively the total magnetization and staggered magnetization which are coarse-grained within an effective cell volume $v_0$. Here, $\epsilon _0$ is the magnetic energy density, and $\mathfrak{h}_0 = \epsilon _0 v_0/(\gamma \hbar)$ is the unit of the magnetic field with $\gamma$ and $\hbar$ being the gyromagnetic ratio and Planck constant, respectively. The distance from the N\'{e}el temperature $T_\text{N}$ is measured by $u_2=(T-T_\text{N})/T_{\text{N}}$, $u_4$ is the quartic coefficient, and $K_0$ ($K_1$) is the uniaxial (in-plane) anisotropy \cite{Yung-Li1964}. The dimensionless coefficient $r_0^{-1}=A/(T+\Theta)$ with two parameters $A$ and $\Theta$ gives the PM susceptibility in the PM phase, and terms proportional to $D$ and $D'$ describe the coupling between $\bm{m}$ and $\bm{n}$. Here and hereafter, we only consider a spatially uniform magnetization dynamics \cite{Yamamoto2019, Brown1963}.
The external magnetic field $\bm{H}_{\text{ex}}$ is given by
\begin{eqnarray}
  \bm{H}_{\text{ex}} = \bm{H}_0 + \bm{h}_{\text{ac}},
\end{eqnarray}
where $\bm{H}_0 = H_0 {\bf \hat{z}}$ is a static magnetic field and $\bm{h}_{\text{ac}}$ is an ac magnetic field, the latter of which is needed to discuss the AFMR modes in the next section.

%
%
%

Next, we focus on the interaction between the AFI and M at the interface. In this work we examine two types of the interfacial exchange coupling. The first is the coupling between the magnetization $\bmm$ in the AFI and the conduction-electron spin density $\bms$ in the M, which we term ``magnetic coupling.'' The second is the coupling between the staggered magnetization $\bmn$ in the AFI and the conduction-electron spin density $\bms$ in the M, which we term ``N\'{e}el coupling.'' Therefore, the interaction between the AFI and M at the interface is described by a free energy, 
\begin{subequations}
  \begin{empheq}[left =
      {   F_{\text{AFI-M}} = \empheqlbrace \,}]{alignat = 2}
    & -J_m \bm{s} \cdot \bm{m}  &\;\;\;&   \text{for \quad magnetic~coupling},
    \label{eq:couplingA}\\
    & -J_n \bm{s} \cdot \bm{n}  & &  \text{for \quad N\'{e}el~coupling},
    \label{eq:couplingB}
    \end{empheq}
\end{subequations}
where $J_m$ ($J_n$) is the magnetic (N\'{e}el) coupling constant at the AFI/M interface. Note that the above form of the exchange interaction at the AFI/M interface is invoked from the knowledge of the exchange bias effect; see Eq.~(28) of Ref.~\cite{Stamps2000}. To the best of our knowledge no quantitative estimate of these coupling constants has been reported so far, which is left for future research.

We now discuss the magnetization dynamics in this bilayer (Fig.~\ref{fig:bilayer}). The dynamics of $\bm{m}$ and $\bm{n}$ in AFI is described by the TDGL equation \cite{Yamamoto2019, Freedman1976, Halperin1976}, 
\begin{eqnarray}
  \label{eq:TDGL_m}
  \frac{\partial}{\partial t} \bm{m}
  &=& \gamma \bm{H}_m \times \bm{m}+\gamma \bm{H}_n \times \bm{n}+\Gamma _m \bm{H}_m/\mathfrak{h}_0 + \bm{\xi},
\\
  \label{eq:TDGL_n}
    \frac{\partial}{\partial t} {\bm n}
  &=& \gamma {\bm H}_n \times {\bm m}+\gamma {\bm H}_m \times {\bm n}+\Gamma _n {\bm H}_n/\mathfrak{h}_0  + \bm{\eta},
\end{eqnarray}
where $\Gamma _m$ and $\Gamma_n$ are damping coefficients for $\bmm$ and $\bmn$, respectively, and the effective fields $\bm{H}_m$ and $\bm{H}_n$ are defined by
\begin{equation}
\bm{H}_m=-\frac{1}{\gamma \hbar} \frac{\partial}{\partial \bm{m}}\big( F_{\text{AFI}}+F_{\text{AFI-M}} \big),
\end{equation}
and
\begin{equation}
\bm{H}_n=-\frac{1}{\gamma \hbar} \frac{\partial}{\partial \bm{n}}\big( F_{\text{AFI}}+F_{\text{AFI-M}} \big).
\end{equation}
The two fields $\bm{\xi}$ and $\bm{\eta}$ in Eqs.~(\ref{eq:TDGL_m}) and (\ref{eq:TDGL_n}) represent thermal noises for $\bm{m}$ and $\bm{n}$, taking the form of Gaussian white noises with zero means and variances:
\begin{equation}
  \average{\xi^i(t)\xi^j(t')}=\frac{2k_{\text{B}} T \Gamma _m}{\epsilon_0 v_0}  \delta_{i,j}\delta(t-t'),
\end{equation}
and
\begin{equation}
  \average{\eta^i(t)\eta^j(t')}=\frac{2k_{\text{B}} T \Gamma _n}{\epsilon_0 v_0}  \delta_{i,j}\delta(t-t'),
\end{equation}
where $i,~j$ denotes $x,~y,~z$, and the noise fields satisfy
\begin{equation}
  \average{\xi^i(t) \eta^j(t')} = 0,
\end{equation}
meaning that $\bm{\xi}$ and $\bm{\eta}$ are statistically independent.

Turning to the M layer, the dynamics of $\bms$ is described by the following TDGL equation:
\begin{equation}
  \label{eq:TDGL_s}
    \frac{\partial}{\partial t} {\bm s}
    = \gamma \bm{H}_s \times \bm{s} + \frac{\chi_{\text{M}} \gamma \hbar}{\tau_{\text{M}}} \bm{H}_s + \bm{\zeta},
\end{equation}
where $\bm{s}$ is the conduction-electron spin density in the M,
\begin{equation}
\bm{H}_s=-\frac{1}{\gamma \hbar} \frac{\partial}{\partial \bm{s}}\big( F_{\text{M}}+F_{\text{AFI-M}} \big)
\end{equation}
is the effective field for $\bms$, and
\begin{equation}
  \label{eq:Free-Energy-M}
  F_{\text{M}}=\frac{1}{2\chi_{\text{M}}}\bm{s}^2
\end{equation}
is the free energy for the M, with $\chi_{\text{M}}$ and $\tau_{\text{M}}$ being the spin susceptibility and spin-flip relaxation time in the M, respectively. The last term on the right-hand side of Eq.~(\ref{eq:TDGL_s}) is the thermal noise field in the M layer, which is represented by a Gaussian white noise with zero mean and variance
\begin{equation}
  \average{\zeta^i(t)\zeta^j(t')}=\frac{2k_{\text{B}} (T+\Delta T) \chi_{\text{M}}}{\tau_{\text{M}}} \delta_{i,j}\delta(t-t').
\end{equation}
Note that the TDGL equation (\ref{eq:TDGL_s}) coincides with the stochastic Bloch equation used to describe the M layer in Ref.~\cite{Yamamoto2019}, where the AF SSE {\it below} the SF transition was investigated. 

In our numerical simulation, we use $\tau_\text{M}$ as the unit of time, $T_{\text{N}}$ as the unit of temperature, $\mathfrak{h}_0$ as the unit of magnetic field, and $\epsilon_0 v_0$ as the unit of energy. Then, we introduce the following rescaling:
\begin{equation}
  \left.
  \begin{array}{ccc}
    t / \tau_{\rm M} & \to & t, \\
    T/T_{\rm N} & \to & T, \\
    \bmH_\text{ex}/\mathfrak{h}_0  & \to & \bmH_\text{ex}.
\end{array}
\right.
\end{equation}
With these dimensionless quantities, the TDGL equations for $\bmm$ and $\bmn$ [Eqs.~(\ref{eq:TDGL_m}) and (\ref{eq:TDGL_n})] read
\begin{eqnarray}
  \label{eq:TDGL_m02}
  \frac{\partial}{\partial t} \bm{m}
  &=& \widetilde{\omega}_0 \bm{H}_m \times \bm{m}
  +\widetilde{\omega}_0 \bm{H}_n \times \bm{n}
  + \widetilde{\Gamma}_m \bm{H}_m + \widetilde{\bm{\xi}},
\\
\label{eq:TDGL_n02}
\frac{\partial}{\partial t} {\bm n}
&=& \widetilde{\omega}_0  {\bm H}_n \times {\bm m}
+ \widetilde{\omega}_0 {\bm H}_m \times {\bm n}
+ \widetilde{\Gamma}_n {\bm H}_n + \widetilde{\bm{\eta}},
\end{eqnarray}
where $\widetilde{\Gamma}_m= \Gamma_m \tau_{\rm M}$, $\widetilde{\Gamma}_n= \Gamma_n \tau_{\rm M}$, and $\widetilde{\omega}_0= \gamma \mathfrak{h}_0 \tau_{\rm M}$. In these equations, $\widetilde{\bm{\xi}}$ and $\widetilde{\bm{\eta}}$ are the dimensionless noise fields with the correlators
\begin{eqnarray}
  \langle \widetilde{\xi}^i (t) \widetilde{\xi}^j (t') \rangle &=& \frac{2 \widetilde{\Gamma}_m \kB T_{\rm N}}{\epsilon_0 v_0} T
  \delta_{i,j} \delta(t-t'), \\
  \langle \widetilde{\eta}^i (t) \widetilde{\eta}^j (t') \rangle &=& \frac{2 \widetilde{\Gamma}_n \kB T_{\rm N}}{\epsilon_0 v_0} T
  \delta_{i,j} \delta(t-t').
\end{eqnarray}
In a similar manner, the TDGL equation for $\bms$ [Eq.~(\ref{eq:TDGL_s})] becomes
\begin{equation}
  \label{eq:TDGL_s02}
    \frac{\partial}{\partial t} {\bm s}
    = \widetilde{\omega}_0 \bm{H}_s \times \bm{s}
    + \widetilde{\chi}_{\text{M}} \bm{H}_s + \widetilde{\bm{\zeta}},
\end{equation}
where $\widetilde{\chi}_{\rm M} = \chi_{\rm M} \epsilon_0 v_0$, and the dimensionless noise field $\widetilde{\bm{\zeta}}$ satisfies 
\begin{equation}
  \langle \widetilde{\zeta}^i (t) \widetilde{\zeta}^j (t') \rangle
  = \frac{2 \widetilde{\chi}_{\rm M} \kB T_{\rm N}}{\epsilon_0 v_0} (T+ \Delta T) 
  \delta_{i,j} \delta(t-t').
\end{equation}
In these equations, the dimensionless interfacial exchange coupling parameter, $\widetilde{J}_m= J_m \tau_{\rm M}/\hbar$ or $\widetilde{J}_n= J_n \tau_{\rm M}/\hbar$, implicitly appears.

To numerically simulate the dimensionless version of the coupled TDGL equations [Eqs.~(\ref{eq:TDGL_m02}), (\ref{eq:TDGL_n02}), and (\ref{eq:TDGL_s02})], we discretize the dimensionless time as ${t}_p= p \Delta t $ ($p= 1, 2, \cdots$)~\cite{Parisi1981}. The resultant discretized version of Eqs.~(\ref{eq:TDGL_m02}), (\ref{eq:TDGL_n02}), and (\ref{eq:TDGL_s02}) is integrated by using the Heun method~\cite{Greiner1988}, where the noises are discretized as $\widetilde{\bm \xi} (t_p)= \int_{t_p}^{t_{p+1}} \widetilde{\bm \xi}(t) dt /\sqrt{\Delta t}$, $\widetilde{\bm \eta}(t_p)= \int_{t_p}^{t_{p+1}} \widetilde{\bm \eta}(t) d t/\sqrt{\Delta t}$, and $\widetilde{\bm \zeta}(t_p)= \int_{t_p}^{t_{p+1}} \widetilde{\bm \zeta}(t) d t/\sqrt{\Delta t}$. Then, the average value of a physical quantity $O$ is obtained by a long-time average over successive measurements from the simulation,
\begin{align}
  \langle O \rangle
  =\frac{1}{N_\text{step}} \sum_{p=1}^{N_\text{step}} O(t_p),
\end{align}
where $N_\text{step}$ is the number of total time steps.

\section{\label{sec:level3}PHASE DIAGRAM AND ANTIFERROMAGNETIC RESONANCE MODES}
In this section, we show the equilibrium phase diagram and AFMR modes calculated from our TDGL model in order to check if the model appropriately describes the known physical properties of uniaxial antiferromagnets. The Ginzburg-Landau approach to the phase diagram including the SF transition can be found in a textbook~\cite{Landau1984}, while the AFMR modes and the AF spin pumping phenomenon have been investigated in recent experiments~\cite{Li2020, Vaidya2020}. As we argue below, the consistency between our result and the properties known in the literature ensures that our numerical calculation correctly captures the low-energy physics of uniaxial antiferromagnets that is relevant to the AF SSE.

\subsection{\label{subsec:level1} Phase diagram and equilibrium quantities}
Here we discuss the phase diagram and equilibrium quantities of the AFI. Analytical expressions for phase boundaries in the $(T, H_0)$ phase diagram of the AFI modeled by \Equref{eq:Free_Energy1} are derived in Appendix~\ref{sec:phase-diagram}. The SF field $H_{\rm SF}$, i.e., the first-order phase transition line between the AF phase and the SF phase, is given by
\begin{equation}
  \label{eq:H_SF}
  H_{\mathrm{SF}}
  =
  \sqrt{\frac{K_0}{D}}
  \left[r_0 -\frac{D+2D'}{2u_4}\left(u_2+\frac{D+D'}{D}K_0\right) \right].
\end{equation}
In addition, a second-order phase transition line $H_{\rm AF/PM}$ separating the AF phase and the PM phase is given by
\begin{equation}
  \label{eq:H_AFPM}
  H_{\text{AF/PM}} =  r_0 \sqrt{\frac{-u_2}{D+D'}},
\end{equation}
whereas another second-order phase transition line $H_{\rm SF/PM}$ separating the SF phase and the PM phase is given by
\begin{equation}
  \label{eq:H_SFPM}
  H_{\rm SF/PM} =  r_0 \sqrt{\frac{-(u_2+K_0)}{D'}}.
\end{equation}

\begin{figure}[t]
  \begin{center}  
    \includegraphics[scale=0.45]{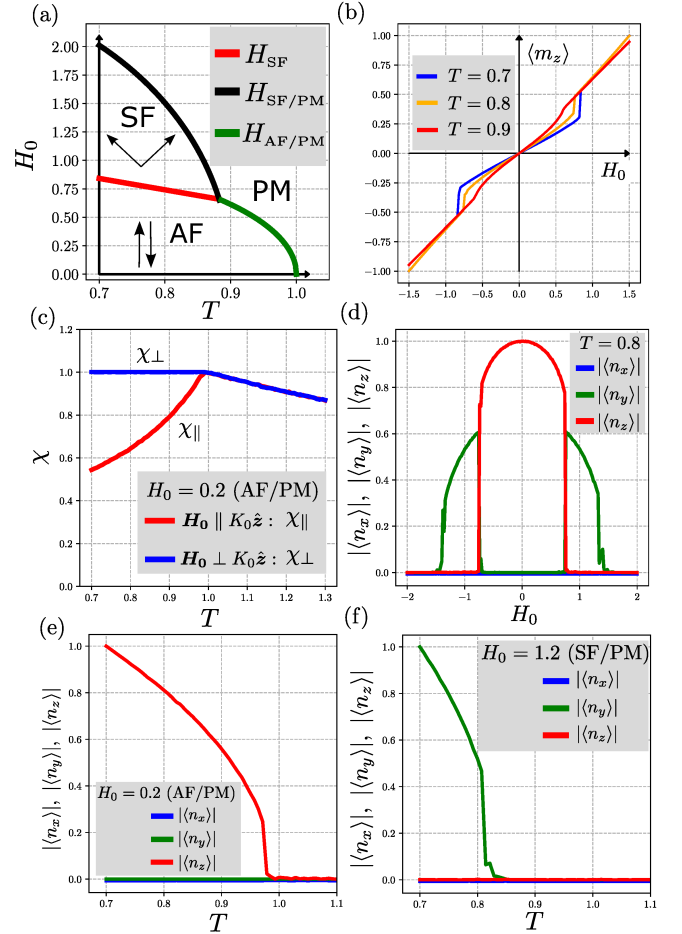}
    \end{center}
\caption{\label{fig:phase_diagram} (a) Phase diagram drawn by Eqs. (\ref{eq:H_SF})--(\ref{eq:H_SFPM}). (b) Magnetic field dependence of the magnetization at several temperatures. (c) Temperature dependence of the spin susceptibility in the AF phase. (d) Magnetic field dependence of the staggered magnetization at $T=0.8$. (e) Temperature dependence of the staggered magnetization in the AF phase. (f) Temperature dependence of the staggered magnetization in the SF phase. In all figures, $u_4=0.1$, $K_0=0.1$, $K_1=0.01$, $D=0.4$, $D'=0.07$, $\Theta/T_\text{N}=1.0$, and $A/T_\text{N}=1.43$ are used. In (b)-(f), $\widetilde{\omega}_0=0.04$, $\widetilde{\Gamma}_m=8\times10^{-4}$, $\widetilde{\Gamma}_n=1.6\times10^{-3}$,
  $\kB T_N/\epsilon_0 v_0 = 1.4 \times 10^{-4}$,
  $\Delta t = 0.1$, and $N_{\rm step} = 3\times 10^7$ are used. The data in (b)-(f) are normalized by the maximum value in each figure.
}
\end{figure}

Figure~\ref{fig:phase_diagram}(a) shows the $(T, H_0)$ phase diagram of AFI drawn by Eqs.~(\ref{eq:H_SF})--(\ref{eq:H_SFPM}). The overall behavior is consistent with the experimental phase diagram for $\rm{MnF_2}$~\cite{Shapira1970}, $\rm{FeF_2}$~\cite{King1983}, and thin films of these materials~\cite{Carrico1994}. Here, we follow the terminology of Ref.~\cite{Shapira1970} for each phase (see Fig.~3 therein). Note that, although the PM phase under magnetic fields possesses a field-induced magnetization, such a forced-ferromagnetic phase is the same as the PM phase in that it does not exhibit any magnetic long-range order (see, also, Sec. 48 of Ref.~\cite{Landau1984}). 

%
%
%

In order to check if our TDGL simulation is consistent with the analytical phase diagram shown in \Figref{fig:phase_diagram}(a), we perform a numerical sampling of the magnetization and staggered magnetization. We first discuss the magnetization. Figure~\ref{fig:phase_diagram}(b) shows the magnetic field dependence of $\langle m_z \rangle$ at several different temperatures. At temperatures $T=0.7$ and $0.8$, we can clearly see that the magnetization abruptly jumps at $H_\text{SF}$, signifying the first-order nature of the phase transition at $H_\text{SF}$. By contrast, at $T=0.9$, the abrupt jump of $\langle m_z \rangle$ disappears and changes to a smooth variation at $H_\text{AF/PM}$, reflecting the second-order transition nature of $H_\text{AF/PM}$. The positions of the phase transitions in \Figref{fig:phase_diagram}(b) are in line with \Figref{fig:phase_diagram}(a), and the behaviors of the magnetization are consistent with the experimental results of Refs.~\cite{Seki2015} and \cite{Wu2016}. 

Figure \ref{fig:phase_diagram}(c) shows the temperature dependence of the parallel spin susceptibility $\widetilde{\chi}_\parallel = \langle m_z \rangle/H_0$ under a field $H_0=0.2$ parallel to the easy AF axis (i.e., ${\bm H}_0 \parallel  {\bf \hat{z}}$). In addition, in order to calculate the perpendicular susceptibility $\widetilde{\chi}_\perp = \langle m_x \rangle/H_0$, we re-orient the external magnetic field to a direction perpendicular to the easy axis (i.e., ${\bm H}_0 \parallel {\bf \hat{x}}$; this field configuration is only used here). In the figure, we see a clear deviation of these two susceptibilities below the AF ordering temperature, which reproduces the well-known result for a uniaxial antiferromagnet~\cite{Kittel1986}.

We next discuss the order parameter of AFI, i.e., the staggered magnetization $\langle \bmn \rangle$. In \Figref{fig:phase_diagram}(d), we show the magnetic field dependence of the staggered magnetization at $T= 0.8$. Since the thermal energy has a sizable weight in comparison to the anisotropy energy and hence an over-barrier transition making changes of the sign of $\langle \bmn \rangle$ occurs in our simulation (see Fig.~6 of~\cite{Wernsdorfer1997} and Fig.~2 of~\cite{Garcia-Palacios1998}), we plot the absolute value of the staggered magnetization. At lower fields (i.e., the AF phase), the staggered magnetization exhibits $\langle n_x \rangle = \langle n_y \rangle = 0$ because $\langle \bmn \rangle$ is oriented along the easy AF axis, i.e.,  $\langle \bmn \rangle \parallel {\bf \hat{z}}$. However, upon increasing the magnetic field, $\langle n_z \rangle$ abruptly disappears at $H_\text{SF}$, followed by a sudden growth of $\langle n_y \rangle$ (i.e., the system enters the SF phase). The $y$ component is selected because the $x$ axis is the hard axis due to the in-plane anisotropy constant $K_1$. Upon further increasing the magnetic field, $\langle n_y \rangle$ disappears at $H_\text{SF/PM}$. These results for $\langle \bmn \rangle$ confirm the consistency of our numerical simulation with the analytic phase diagram [\Figref{fig:phase_diagram}(a)] along a line $\text{AF} \rightarrow \text{SF} \rightarrow \text{PM}$ at $T=0.8$. We have also confirmed that along other constant-temperature lines our numerical simulation is always consistent with the phase diagram given in \Figref{fig:phase_diagram}(a). The discontinuous change of the direction of $\langle \bmn \rangle$ at $H_\text{SF}$ signifies that it is the first-order phase transition.

Next, in \Figref{fig:phase_diagram}(e), we show the temperature dependence of the staggered magnetization across the AF/PM phase boundary at $H_0=0.2$. In the figure we clearly see that the staggered magnetization is ordered along the easy AF axis in the AF phase, and it vanishes at $H_\text{AF/PM}$. Similarly, in \Figref{fig:phase_diagram}(f), we show the temperature dependence of $\langle \bmn \rangle$ across the SF/PM phase at $H_0=1.2$. In this case the staggered magnetization is ordered perpendicularly to the easy axis ($\langle \bmn \rangle \perp {\bf \hat{z}}$), and it disappears at $H_\text{SF/PM}$. In both cases, the staggered magnetization continuously vanishes at $H_\text{AF/PM}$ and $H_\text{SF/PM}$, such that these boundaries signify the second-order phase transition.

From these results, we can confirm that our numerical simulation is consistent with the analytic phase diagram given in \Figref{fig:phase_diagram}(a).
%
%
\subsection{\label{subsec:level2}Antiferromagnetic resonance modes}
Next, we discuss the AFMR modes of the AFI. Analytical expressions for the AFMR modes of the AFI modeled by \Equref{eq:Free_Energy1} are derived in Appendix~\ref{sec:AFMRmode}. 

In the AF phase, the resonance modes are given by
\begin{eqnarray}
  \label{eq:AFMR}
  \omega_{\pm}
  &=&
  \frac{1}{2} \widetilde{\omega}_0  H_0
  \big[ 1 + \widetilde{\chi}_{\parallel}\{D(n_{\text{eq}}^2- m_{\text{eq}}^2)+K_0\} \big]
 \nonumber\\
  && \pm \frac{1}{2}
 \Big[(\widetilde{\omega}_0 H_0)^2\big\{ \widetilde{\chi}_{\parallel} \left(D(m_{\text{eq}}^2-n_{\text{eq}}^2)-K_0 \right)+1
   \big\}^2 \nonumber \\
   && + 4(\widetilde{\omega}_0 n_{\text{eq}})^2 K_0( \widetilde{\chi}_{\perp}^{-1}+Dm_{\text{eq}}^2)
   \Big]^{1/2},
\end{eqnarray}
where $\widetilde{\chi}_\parallel=\left( r_0 + (D+D')n_\text{eq}^2 \right)^{-1}$ and $\widetilde{\chi}_\perp=\left( r_0 + D'n_\text{eq}^2 \right)^{-1}$ are the parallel and perpendicular spin susceptibilities in the AF phase. Following the terminology of Ref.~\cite{Li2020}, we call the $+$ branch the ``right-handed'' (RH) mode, whereas the $-$ branch the ``left-handed'' (LH) mode. Note that we define the frequency of the LH mode so as to have the negative sign because, with this definition, it is easier to distinguish the LH mode from the RH one, whereas the conventional definition gives a positive sign for both modes~\cite{Li2020,Vaidya2020}. 

In the SF phase, on the other hand, one resonance frequency is given by

\begin{flalign}
  \label{eq:QFMR}
  \omega _{\mathrm{QFMR}}
  =
  \sqrt{(\widetilde{\omega}_0 H_0)^2 - (\widetilde{\omega}_0 n_{\text{eq}})^2K_0/\widetilde{\chi}_{\text{SF}} },
\end{flalign}
where  $\widetilde{\chi}_\text{SF}=(r_0+D'n_\text{eq}^2)^{-1}$ is the spin susceptibility in the SF phase. Hereafter, this mode is called the ``quasi-ferromagnetic resonance'' (QFMR) mode~\cite{Li2020,Vaidya2020} in this paper. Another resonance frequency in the SF phase is given by 
\begin{equation}
  \label{eq:flat-mode}
  \omega _{\text{flat}}
  =
  \sqrt{K_1 n_{\text{eq}}^2\{\widetilde{\omega}_0^2/\widetilde{\chi}_{\text{SF}}+2(\widetilde{\omega}_0 H_0 \widetilde{\chi}_{\text{SF}})^2(u_4-2D')\}},
\end{equation}
which is hereafter called the ``flat mode''.

\begin{figure}[t]
  \begin{center}
    \includegraphics[scale=0.47]{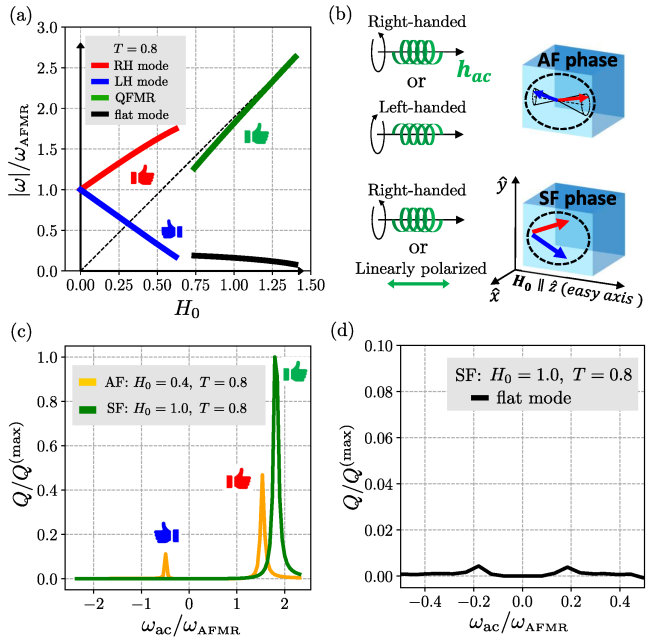}
    \end{center}
\caption{\label{fig:dispersion} (a) Absolute value of the AFMR frequencies [Eqs. (\ref{eq:AFMR})--(\ref{eq:flat-mode})] in the AF and the SF phases as a function of external magnetic field $H_0$. Here the frequency is renormalized by the zero-field resonance frequency $\omega_{\text{AFMR}}$ [\Equref{eq:AFMR00}]. The dashed black line denotes the PM resonance frequency $\widetilde{\omega}_0 H_0$, and the hand symbol signifies the handedness of each mode. 
  (b) Schematic illustration of the AFMR in the AF phase (upper) and the SF phase (lower). The magnon resonates with the RH/LH ac field in the former, whereas it resonates with the RH/linearly polarized ac field in the latter.
  (c) Power absorption in the AF phase (orange) and SF phase (green) at $T=0.8$, obtained in our TDGL simulation for a circularly polarized ac magnetic field [\Equref{eq:hac_clr}]. The power absorption is normalized by the maximum value. (d) Magnified view of the power absorption in the SF phase for a linearly polarized ac magnetic field [\Equref{eq:hac_lnr}]. The normalization of the vertical axis is the same as (c). For details, see the main text.
  Here, $u_4=0.1$, $K_0=0.1$, $K_1=0.01$, $D=0.4$, $D'=0.07$, $\Theta/T_\text{N}=1.0$, $A/T_\text{N}=1.43$, $\widetilde{\omega}_0=0.04$, $\widetilde{\Gamma}_m=8\times10^{-4}$, $\widetilde{\Gamma}_n=1.6\times10^{-3}$, $\Delta t = 0.1$, $T=0.8$, $h_{\mathrm{ac}}=10^{-3}$, $\kB T_N/\epsilon_0 v_0 = 1.4 \times 10^{-4}$, $\widetilde{J}_m = \widetilde{J}_n= 0$, and $N_{\rm step}= 5\times 10^3 $ are used.}
\end{figure}

In \Figref{fig:dispersion}(a), the AFMR frequencies calculated from Eqs. (\ref{eq:AFMR})--(\ref{eq:flat-mode}) are plotted as a function of the magnetic field $H_0$, where the frequency is normalized by the zero-field resonance frequency
\begin{equation}
  \omega_{\text{AFMR}} = \widetilde{\omega}_0 n_{\text{eq}} \sqrt{K_0 / \widetilde{\chi}_{\perp}}.
  \label{eq:AFMR00}
\end{equation}

%
%
%
%

Let us now turn to the TDGL simulation of the AFMR modes. As schematically depicted in \Figref{fig:dispersion}(b), we use the following circularly polarized ac magnetic field to resonate the AFI system:
\begin{eqnarray}
  \bm{h}_{\mathrm{ac}}(t)=h_{\mathrm{ac}}(\cos \omega_{\mathrm{ac}}t,~\sin \omega_{\mathrm{ac}} t,~0),
  \label{eq:hac_clr}
\end{eqnarray}
where we define the positive sign of $\omega_{\mathrm{ac}}$ so as to give the right-handed helicity. In the AFMR, the AFI system establishes a steady state by absorbing and releasing the same amount of energy. The mean power absorbed to the system per unit time can be evaluated by the following formula \cite{Landau1984a}:
\begin{eqnarray}
  \label{eq:heat}
  Q
  =
  -\left[ \langle {\bm{m}} \rangle \cdot \frac{d \bm{h}_{\mathrm{ac}}(t)}{dt} \right]_{T_{\mathrm{ac}}}, \label{eq:Q}
\end{eqnarray}
where $[\cdots]_{T_{\rm ac}}= \frac{1}{T_{\rm ac}} \int_0^{T_{\rm ac}} dt (\cdots)$ denotes the time average over one period of the ac magnetic field, $T_{\text{ac}}=|2\pi/\omega_{\mathrm{ac}}|$. We numerically sample the power absorption [\Equref{eq:Q}] by simulating Eqs.~(\ref{eq:TDGL_m}) and (\ref{eq:TDGL_n}).

Figure~\ref{fig:dispersion}(c) shows the result of our TDGL simulation for the power absorption plotted as a function of $\omega_{\rm ac}$ for a constant magnetic field and $T=0.8$. In the AF phase [orange line in \Figref{fig:dispersion}(c)], we observe two peaks at $\omega_{\text{ac}}/\omega_{\text{AFMR}} \approx -0.5,~1.5$, which can be inferred from \Figref{fig:dispersion}(a) by sweeping $\omega_{\rm ac}$ at $H_0= 0.4$ and finding a cross point with the LH mode (RH mode) at $|\omega_{\text{ac}}|/\omega_{\text{AFMR}} \approx 0.5$ ($|\omega_{\text{ac}}|/\omega_{\text{AFMR}} \approx 1.5$). In the SF phase [green line in \Figref{fig:dispersion}(c)], on the other hand, we observe one peak at $\omega_{\text{ac}}/\omega_{\text{AFMR}} \approx 1.8$ which can also be inferred from \Figref{fig:dispersion}(a) by sweeping $\omega_{\rm ac}$ at $H_0= 1.0$ and finding a cross point with the QFMR mode. Therefore, the resonance positions in our TDGL simulation are in good agreement with the analytical calculation of the AFMR frequency shown in Fig. \ref{fig:dispersion}(a) for the RH, LH, and QFMR modes.

By contrast, confirmation of the flat mode $\omega_\text{flat}$ requires some care, since it has a linear polarization along the $z$ axis~\cite{Reitz2020}. Therefore, instead of the circularly polarized field given by \Equref{eq:hac_clr}, we apply the following linearly polarized field: 
\begin{equation}
  \bm{h}_\text{ac}=h_\text{ac}\left(0, \, 0, \, \cos \omega_\text{ac}t\right).
  \label{eq:hac_lnr}
\end{equation}
Figure~\ref{fig:dispersion}(d) shows the power absorption numerically simulated for an external ac magnetic field given by \Equref{eq:hac_lnr}. Here, we see a very tiny peak around $|\omega_{\text{ac}}|/\omega_{\text{AFMR}} \approx 0.2$. The reason for this weak resonance could be due to the extremely low frequency of the flat mode, which is proportional to the square root of the weak in-plane anisotropy constant $K_1$ [see \Equref{eq:flat-mode}]. 
    
The results in this section demonstrate that our TDGL simulation correctly captures the dynamics of two magnon modes in the AF phase (the RH and LH modes) with opposite helicities, the QFMR mode in the SF phase with the helicity the same as the RH mode, and the flat mode with linear polarization. Therefore, we can confirm that the low-energy excitations needed to describe the AF SSE are well reproduced in our TDGL simulation.

%
%

\section{\label{sec:level4}ANTIFERROMAGNETIC SPIN SEEBECK EFFECT}
We are now in a position to discuss the AF SSE. In this section, we perform a TDGL simulation of the AF SSE to investigate the sign reversal phenomenon across the SF transition. In our simulation, not only the TDGL equations for the AFI layer [Eqs.~(\ref{eq:TDGL_m}) and (\ref{eq:TDGL_n})] but also the TDGL equation for the M layer [Eq.~(\ref{eq:TDGL_s})] are numerically solved with the Heun method~\cite{Greiner1988} because these stochastic equations are coupled through the interfacial exchange interaction [Eqs.~(\ref{eq:couplingA}) and (\ref{eq:couplingB})]. This point, that we directly evaluate the spin current injected into the M layer by tracking the time evolution of $\bms$, distinguishes our approach from the previous works relying on the spin-mixing conductance phenomenology~\cite{Barker2016,Reitz2020}. Note that the numerical cost in our approach becomes more expensive due to the increase in the dynamical variables. The advantage of our approach is that a more microscopic origin of the sign reversal of the AF SSE can be identified. 

The spin current injected into the M layer is defined as the rate of change of the spin density, i.e.,
\begin{equation}
  I_s = \average{\partial _t s_z(t)}.
  \label{eq:Is}
\end{equation}
The right-hand side of this equation can be transformed using the $z$ component of the Bloch equation (\ref{eq:TDGL_s02}). Then, when we discuss the magnetic coupling [\Equref{eq:couplingA}], the spin current is given by
\begin{eqnarray}
  I_s &=&
  \widetilde{J}_m
  \langle
  m_x s_y - s_x m_y
  \rangle,
\end{eqnarray}
whereas when we discuss the N\'{e}el coupling [\Equref{eq:couplingB}], it becomes
\begin{eqnarray}
    I_s &=&
  \widetilde{J}_n
  \langle
  n_x s_y - s_x n_y
  \rangle.
\end{eqnarray}
In addition to the interfacial exchange coupling parameters $J_m$ and $J_n$ characterizing two types of the interfacial exchange interaction, another important parameter is the dimensionless parameter

\begin{equation}
  \widetilde{\omega}_0=\gamma \mathfrak{h}_0 \tau_\text{M},
\end{equation}
which leads to two different behaviors in our simulation. Physically, this parameter measures the strength of the spin dephasing in the M layer. Below, we present results of our simulations for two different values of $\widetilde{\omega}_0= 0.5$ and $\widetilde{\omega}_0= 0.04$.

%
\begin{figure}[t]
  \begin{center}
    \includegraphics[scale=0.58]{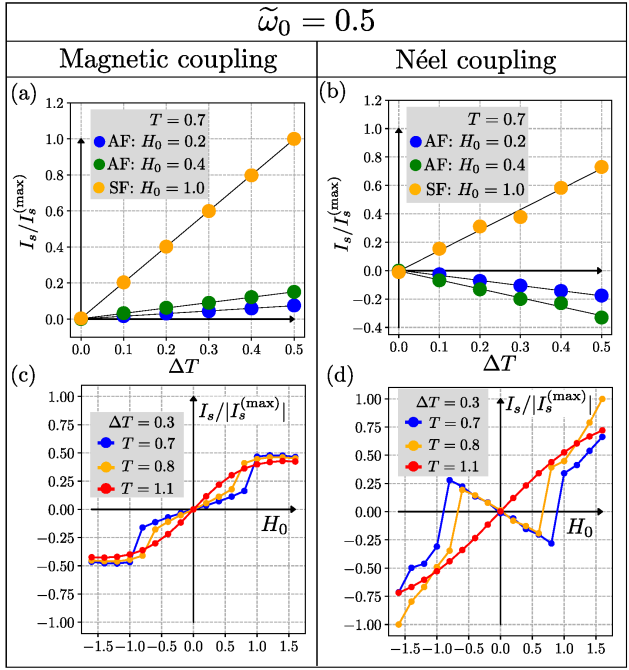}
  \end{center}
\caption{\label{fig:omega0_0p50}
  Spin current obtained for the (a),(c) magnetic coupling and (b),(d) N\'{e}el coupling using $\widetilde{\omega}_0=0.5$. In (a) and (b), the spin current at $T=0.7$ is plotted as a function of temperature bias $\Delta T$ for several values of $H_0$, where the black solid line shows a linear regression fit. In (c) and (d), the magnetic field dependence of the spin current is plotted for several temperatures. In (a) and (b), the data are normalized by the maximum value in (a) in order to enable the comparison of (a) and (b), whereas in (c) and (d), the data are normalized by the maximum value in (d). Here, we use $\widetilde{\Gamma}_m=0.01$, $\widetilde{\Gamma}_n=0.02$, $\widetilde{J}_m = 0.01$  (magnetic coupling), $\widetilde{J}_n=0.01$ (N\'{e}el coupling), and $\kB T_N/\epsilon_0 v_0 = 1.4 \times 10^{-4}$. $N_\text{step}= 1\times10^{10}$ in (a) and (c), while $1\times10^{11}$ in (b) and (d).
  Other parameters characterizing the equilibrium properties are the same as in \Figref{fig:phase_diagram}, i.e.,
  $u_4=0.1$, $K_0=0.1$, $K_1=0.01$, $D=0.4$, $D'=0.07$, $\Theta/T_\text{N}=1.0$, and $A/T_\text{N}=1.43$.
}
\end{figure}
%
%

%
%
\begin{figure}[t]
  \begin{center}
    \includegraphics[scale=0.58]{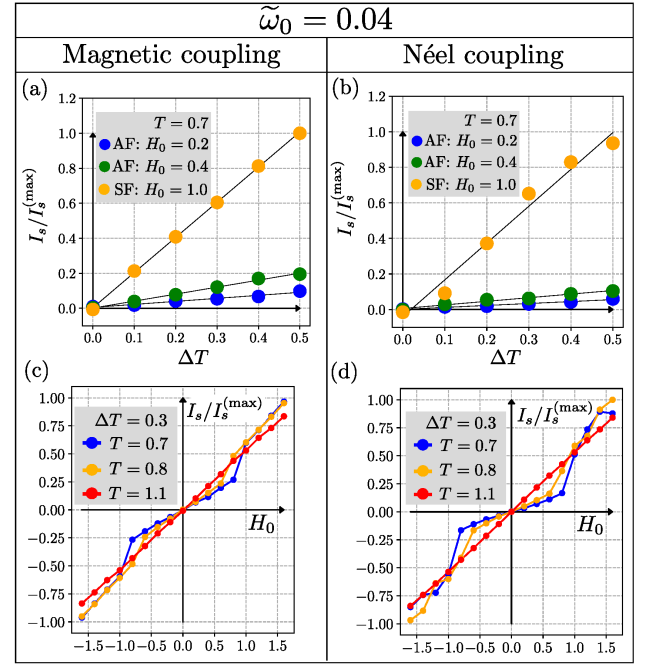}
  \end{center}    
\caption{\label{fig:omega0_0p04}
  Same as Fig.~\ref{fig:omega0_0p50} but using $\widetilde{\omega}_0=0.04$. Here, we use  $\widetilde{\Gamma}_m=8.0\times10^{-4}$, $\widetilde{\Gamma}_n=1.6\times10^{-3}$, $\widetilde{J}_m= 8.0\times10^{-4}$ (magnetic coupling), $\widetilde{J}_n=8.0\times10^{-4}$ (N\'{e}el coupling), and 
$\kB T_N/\epsilon_0 v_0 = 1.4 \times 10^{-4}$. (a),(c) $N_\text{step} = 3\times 10^{10}$, (b) $8\times10^{11}$, and (d) $2\times10^{12}$. 
}
\end{figure}
%
%

Figure~\ref{fig:omega0_0p50} shows the spin current obtained by a TDGL simulation for $\widetilde{\omega}_0= 0.5$. In Figs.~\ref{fig:omega0_0p50}(a) and \ref{fig:omega0_0p50}(b), the spin current is plotted as a function of temperature bias $\Delta T$ for the magnetic coupling [Fig.~\ref{fig:omega0_0p50}(a)] and N\'{e}el coupling [Fig.~\ref{fig:omega0_0p50}(b)], respectively. The simulations are done for both the AF phase and the SF phase [see Fig.~\ref{fig:phase_diagram}(a) for the phase diagram]. As we can see, in both phases the spin current is proportional to the temperature bias with zero intercept, demonstrating that our simulation is within the linear response regime as in experiments~\cite{Uchida2010a}. Note that the sign of the spin current for the magnetic coupling [Fig.~\ref{fig:omega0_0p50}(a)] is always positive, whereas its sign for the N\'{e}el coupling [Fig.~\ref{fig:omega0_0p50}(b)] becomes negative in the AF phase. In Figs.~\ref{fig:omega0_0p50}(c) and \ref{fig:omega0_0p50}(d), magnetic field dependence of the spin current is plotted for the magnetic coupling [Fig.~\ref{fig:omega0_0p50}(c)] and N\'{e}el coupling [Fig.~\ref{fig:omega0_0p50}(d)], respectively. In the case of the magnetic coupling [\Figref{fig:omega0_0p50}(c)], for $T= 0.7$ and $0.8$, the SSE signal grows with positive sign at lower $H_0$, and there is an abrupt jump in the SSE signal with no sign reversal across the SF transition. By contrast, in the case of the N\'{e}el coupling [\Figref{fig:omega0_0p50}(d)], for $T= 0.7$ and $0.8$, the SSE signal starts from a negative sign at lower $H_0$, and it suddenly reverses its sign across the SF transition. Finally, turning to the spin current in the PM region, the SSE signal is always positive and there is no singular behavior in the signal. Our result shown in Fig~\ref{fig:omega0_0p50}(c) is consistent with the experiments reported in Refs.~\cite{Seki2015} and \cite{Wu2016}, whereas the one shown in Fig~\ref{fig:omega0_0p50}(d) reproduces the experimental result of Refs.~\cite{Li2020} and \cite{Yuan2020}.

Next, we demonstrate that the sign reversal of the AF SSE across the SF transition [Fig.~\ref{fig:omega0_0p50}(d)] disappears for a too strong spin dephasing in the M layer (i.e., for smaller $\widetilde{\omega}_0$). Figure~\ref{fig:omega0_0p04} shows the spin current simulated for $\widetilde{\omega}_0= 0.04$. In Figs.~\ref{fig:omega0_0p04}(a) and \ref{fig:omega0_0p04}(b), the spin current is plotted as a function of temperature bias $\Delta T$ for the magnetic coupling [Fig~\ref{fig:omega0_0p04}(a)] and N\'{e}el coupling [Fig~\ref{fig:omega0_0p04}(b)], respectively. As in Figs.~\ref{fig:omega0_0p50}(a) and \ref{fig:omega0_0p50}(b), the TDGL simulation results belong to the linear response regime. In Figs.~\ref{fig:omega0_0p04}(c) and \ref{fig:omega0_0p04}(d), the magnetic field dependence of the spin current is plotted for the magnetic coupling [Fig.~\ref{fig:omega0_0p04}(c)] and N\'{e}el coupling [Fig.~\ref{fig:omega0_0p04}(d)], respectively. A crucial difference from Figs.~\ref{fig:omega0_0p50}(c) and \ref{fig:omega0_0p50}(d) is that no sign reversal can be found in Figs.~\ref{fig:omega0_0p04}(c) and \ref{fig:omega0_0p04}(d). Since $\widetilde{\omega}_0$ is the only the parameter changed from Fig.~\ref{fig:omega0_0p50}, the result shows that the stronger spin dephasing (smaller $\widetilde{\omega}_0$) makes the sign reversal vanish.

\begin{figure}[t]
  \begin{center}
    \includegraphics[scale=0.49]{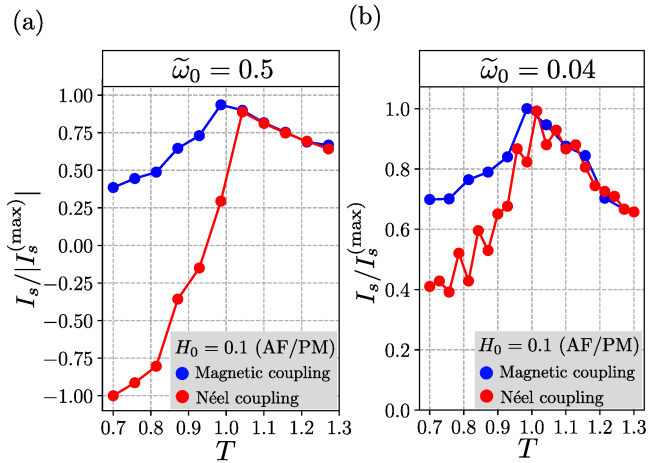}
    \end{center}
\caption{\label{fig:SSE-Tdepend}
  Temperature dependence of the spin current for (a) $\widetilde{\omega}_0=0.5$ and (b) $\widetilde{\omega}_0=0.04$. The temperature bias is fixed at $\Delta T = 0.3$. In (a), we use $N_{\text{step}}=1\times 10^{10}$ for the magnetic coupling and $N_{\text{step}}=3\times 10^{11}$ for the N\'{e}el coupling, and the other parameters are the same as \Figref{fig:omega0_0p50}. In (b), we use $N_{\text{step}}=3\times 10^{11}$ for the magnetic coupling, and $N_{\text{step}}=2\times 10^{12}$ for the N\'{e}el coupling. The other parameters are the same as \Figref{fig:omega0_0p04}. The data are normalized by the maximum value of $|I_s^{\rm (max)}|$ in each figure.
}
\end{figure}

Finally, we examine the temperature dependence of the AF SSE. Figure \ref{fig:SSE-Tdepend} shows the spin current as a function of temperature at $H_0= 0.1$ for $\widetilde{\omega}_0=0.5$ [Fig.~\ref{fig:SSE-Tdepend}(a)] and $\widetilde{\omega}_0=0.04$ [Fig.~\ref{fig:SSE-Tdepend}(b)]. As concluded in Ref.~\cite{Yamamoto2019}, we see that there appears a cusp at $T=T_{\rm N}$ in the signal, reflecting that the AF SSE is proportional to the spin susceptibility which possesses the cusp structure at the N\'{e}el temperature. We also see that, in the case of the N\'{e}el coupling, the AF SSE reverses its sign below the N\'{e}el temperature for $\widetilde{\omega}_0=0.5$, but this sign reversal disappears for $\widetilde{\omega}_0=0.04$.

The result of our TDGL simulation in this section demonstrates that the presence of a sizable N\'{e}el coupling is the origin of the sign reversal in the AF SSE across the SF transition. Moreover, it also demonstrates that the sign reversal disappears for a too strong spin dephasing in the M layer.

%
%
%
%
%
\section{\label{sec:level5} Discussion of the results}
In this section, we discuss the results obtained in the preceding section by focusing on the condition for the sign reversal of the AF SSE across the SF transition. We first derive an analytical expression for the AF SSE in the AF phase by extending the calculation of our previous work~\cite{Yamamoto2019}, and then discuss the condition for the sign reversal. For this purpose, we use the dimensionful form of the TDGL equations (\ref{eq:TDGL_m}), (\ref{eq:TDGL_n}), and (\ref{eq:TDGL_s}), instead of their dimensionless versions [Eqs. (\ref{eq:TDGL_m02}), (\ref{eq:TDGL_n02}), and (\ref{eq:TDGL_s02})]. 

We first note that in \Figref{fig:omega0_0p50}(d), a {\it negative} sign of the AF SSE signal is obtained in the AF phase, where an analytical calculation of the AF SSE is possible along the line of calculation in Ref.~\cite{Yamamoto2019}. Here, we extend that calculation~\cite{Yamamoto2019} to the one including higher-order terms with respect to $\omega_\pm \tau_{\rm M}$, where $\omega_\pm$ is the RH/LH mode frequency defined by \Equref{eq:AFMR}. Then, in the case of the magnetic coupling, the spin current driven by the AF SSE is given by (see Appendix~\ref{sec:analytical_spin_current} for details)
\begin{equation}
  \label{eq:I_m}
  I_s
  =
  J_m^2 {\cal K} \Delta T
  \frac{  1  -  |\omega_{+}\omega_{-}|\tau_\text{M}^{2}  K_0 \widetilde{\chi}_{\parallel}  }
       {   (1+ \omega_{+}^2 \tau_\text{M}^{2})(1+ \omega_{-}^2 \tau_\text{M}^{2})   },
\end{equation}
whereas in the case of the N\'{e}el coupling, it becomes
\begin{equation}
  \label{eq:I_n}
  I_s
  =
  J_n^2 {\cal K} \Delta T
  \frac{ \;\; 1 - |\omega_{+}\omega_{-}|\tau_\text{M}^{2}(K_0 \widetilde{\chi}_{\parallel})^{-1} }
       { (1+ \omega_{+}^2 \tau_\text{M}^{2})(1+ \omega_{-}^2 \tau_\text{M}^{2}) },
\end{equation}
where ${\cal K}= {2k_\text{B} \chi_\text{M} m_{\text{eq}}\tau_\text{M}}/\hbar^2$ is a positive coefficient, and $m_{\text{eq}}= \widetilde{\chi}_{\parallel}H_0/\mathfrak{h}_0$ is the field-induced magnetization in the AFI under a magnetic field $H_0 {\bf \hat{z}}$ along the anisotropy axis. Note that in the limit of $|\omega_\pm| \tau_{\rm M} \ll 1$, the above results reduce to the result of Ref.~\cite{Yamamoto2019}. Note also that as emphasized below \Equref{eq:AFMR}, the LH mode $\omega_-$ has a negative sign in our definition.

In discussing the above two Eqs. (\ref{eq:I_m}) and (\ref{eq:I_n}), it is important to recall that the factor $K_0 \widetilde{\chi}_{\parallel}$ is small in realistic situations, i.e., $K_0 \widetilde{\chi}_{\parallel} \ll 1$, and that a condition $|\omega_+ \omega_-| \tau_{\rm M}^2 \ll 1$ holds in order for the M layer to act as a good spin sink. With these conditions, we see that the AF SSE that emerges from the magnetic coupling [\Equref{eq:I_m}] is always positive for $\Delta T > 0$~\cite{Yamamoto2019} (note that the definition of positive $\Delta T$ is opposite to Ref.~\cite{Yamamoto2019}). By contrast, the AF SSE that emerges from the N\'{e}el coupling [\Equref{eq:I_n}] can be negative when the condition
\begin{equation}
  \label{eq:reversal-condition}
  |\omega_{+} \omega_{-}| \tau_\text{M}^2 \gtrsim {K_0 \widetilde{\chi}_{\parallel}}
\end{equation}
is satisfied, which in the dimensionless form used in the numerical simulation, is transformed into
\begin{equation}
  \label{eq:reversal-condition2}
  |\widetilde{\omega}_{+} \widetilde{\omega}_{-}| \widetilde{\omega}_0^2
  \gtrsim {K_0 \widetilde{\chi}_{\parallel}},
\end{equation}
where we introduced $\widetilde{\omega}_\pm = \omega_\pm /\gamma \mathfrak{h}_0$.  In \Equref{eq:reversal-condition2}, the only parameter that depends on a material choice for the M layer is $\widetilde{\omega}_0$. Therefore, a larger $\widetilde{\omega}_0$ allows \Equref{eq:reversal-condition2} to be satisfied. This is indeed seen in \Figref{fig:SSE-Tdepend}, where the sign reversal of the N\'{e}el spin current occurs for a larger $\widetilde{\omega}_0= 0.5$ [\Figref{fig:SSE-Tdepend}(a)], whereas such a sign reversal does not happen for a smaller $\widetilde{\omega}_0= 0.04$ [\Figref{fig:SSE-Tdepend}(b)].

Before proceeding, it is informative to describe the difference between the present formalism using the interfacial exchange coupling~\cite{Yamamoto2019} and the one using the spin-mixing conductance~\cite{Reitz2020}. According to the spin-mixing conductance formalism~\cite{Reitz2020}, the spin pumping current is given by 
\begin{equation}
  I_s^{\rm pump} = {g}_m {\bf \hat{z}}\cdot \langle \bm{m} \times \partial _t \bm{m} \rangle
  \label{eq:Ispump_m}
\end{equation}
in the case of the magnetic coupling, and 
\begin{equation}
  I_s^{\rm pump} = {g}_n {\bf \hat{z}} \cdot \langle \bm{n} \times \partial _t \bm{n} \rangle
  \label{eq:Ispump_n}
\end{equation}
in the case of the N\'{e}el coupling, where $g_m$ and $g_n$ are the corresponding spin-mixing conductances. Then, the AF SSE signal can be expressed as 
  \begin{equation}
    I_s = I_s^{\rm pump}  - I_s^{\rm back}, 
  \end{equation}
  where $I_s^{\rm back}$ is the backflow current. Note that the back flow current is related to the pumping current by $I_s^{\rm back}(T)= I_s^{\rm pump}(T+\Delta T)$ due to the fluctuation-dissipation relation~\cite{Foros2005}. As discussed in Appendix~\ref{sec:analytical_spin_current}, these spin-mixing conductances are obtained within our formalism as $g_m= J_m^2 \chi_{\rm M} \tau_{\rm M}/\hbar$ and $g_n= J_n^2 \chi_{\rm M} \tau_{\rm M}/\hbar$, where this correspondence holds only in the $|\omega_\pm| \tau_{\rm M} \to 0$ limit. Therefore, the difference between the two formalisms is that our formalism can take into account the dynamical information of the M layer as well as the AFI layer, but the spin-mixing conductance formalism deals only with the dynamics of the AFI layer.

A physical interpretation of Eqs.~(\ref{eq:I_m}) and (\ref{eq:I_n}) regarding the sign of the AF SSE is as follows. First, it is important to note that the RH mode has a positive helicity with higher frequencies, whereas the LH mode has a negative helicity with lower frequencies [see Fig.~\ref{fig:dispersion}(a)]. Second, a higher frequency of the RH magnon results in a smaller magnon population $\sim T/\omega$, but at the same time it results in a larger pumping current since the latter is proportional to the magnon frequency, i.e., $I_s^{\rm pump} \propto \omega {\bf \hat{z}} \cdot \langle \bmm_\omega \times \bmm^*_\omega \rangle$ for the magnetic coupling, and $I_s^{\rm pump} \propto \omega {\bf \hat{z}} \cdot \langle \bmn_\omega \times \bmn^*_\omega \rangle$ for the N\'{e}el coupling. Then, within the present TDGL approach to the AF SSE valid at $\kB T \gg \hbar \omega$, the former disadvantage of the RH magnon with a smaller population is overwhelmed by the latter advantage with a larger pumping current. Namely, in both cases of the magnetic and N\'{e}el couplings, the first term in the numerator of Eqs.~(\ref{eq:I_m}) and (\ref{eq:I_n}) is determined by the RH mode magnons, thereby giving the positive sign of the AF SSE. The difference between the magnetic and N\'{e}el couplings appears in the second term in the numerator of Eqs.~(\ref{eq:I_m}) and (\ref{eq:I_n}). Because this term is higher order with respect to $|\omega_+ \omega_- |\tau_{\rm M}^2$ that involves the timescale of the M, its effect is considered to come from the backflow current $I_s^{\rm back}$ that produces a negative contribution. In the case of the magnetic coupling, the correction to the backflow current is so small, of the order of $(K_0 \widetilde{\chi}_{\parallel}) \ll 1$, and therefore no sign reversal happens. By contrast, in the case of the N\'{e}el coupling the correction to the backflow current is huge, of the order of $(K_0 \widetilde{\chi}_{\parallel})^{-1} \gg 1$. This brings about a sign reversal of the AF SSE when the condition of \Equref{eq:reversal-condition} is satisfied. 

To summarize this section, the sign reversal of the AF SSE across the SF transition requires the following two conditions: (i) the presence of a sizable N\'{e}el coupling $J_n$ [\Equref{eq:I_n}] and (ii) moderate strength of the spin dephasing $\tau_{\rm M}^{-1}$ in the M layer [\Equref{eq:reversal-condition}].

%
%
\section{\label{sec:level6}CONCLUSION}

The main result of this paper is that under certain conditions the sign of the AF SSE is reversed across the SF transition, but the sign reversal is not the generic property of a simple uniaxial antiferromagnet. Indeed, the sign reversal requires the following two conditions: (i) there exists a sizable N\'{e}el coupling $J_n$ at the AFI/M interface that produces the N\'{e}el spin current given by \Equref{eq:I_n}, and (ii) the spin dephasing $\tau_{\rm M}^{-1}$ of the M layer is not too strong so that it satisfies the inequality given by \Equref{eq:reversal-condition}.

Although the sign reversal of the AF SSE across the SF transition has been discussed by Reitz \textit{et al.}~\cite{Reitz2020} theoretically, our approach has several advantages over their phenomenological one~\cite{Reitz2020}. First, the present approach has identified that the N\'{e}el current [\Equref{eq:I_n}] producing the sign reversal originates microscopically from the N\'{e}el coupling given by \Equref{eq:couplingB}. A possible way of experimentally realizing the N\'{e}el coupling is to prepare an atomically sharp uncompensated interface~\cite{Daniels2015}. Also, since the N\'{e}el coupling has the same symmetry as the exchange bias effect in a ferromagnet/antiferromagnet bilayer~\cite{Coey2009} under the sublattice exchange transformation, we expect that the tendency for an antiferromagnet to exhibit the exchange bias effect is the key parameter controlling the sign reversal. Second, we provide a possible scenario for the {\it absence} of the sign reversal in previous experiments for $\rm{Cr_2O_3}$~\cite{Seki2015} and $\rm{MnF_2}$~\cite{Wu2016}, as well as its {\it presence} in recent experiments for $\rm{Cr_2O_3}$~\cite{Li2020} and $\alpha$-$\rm{Fe_2O_3}$~\cite{Yuan2020}, once we are allowed to assume that the details of the interfacial exchange coupling depend on material choices and sample fabrication conditions. Needless to say, the validity of our scenario should be examined by future experiments. Third, the present approach allows us to calculate temperature dependence of the AF SSE. In particular, our result may account for the disappearance of the sign reversal as a function of temperature in a $\rm{Cr_2O_3}/{\rm Pt}$ bilayer~\cite{Li2020} (see Fig. 4d therein), as such a disappearance is indeed demonstrated in \Figref{fig:SSE-Tdepend}(a). 

To conclude, we have demonstrated with the stochastic TDGL simulation that a sign reversal of the AF SSE appears across the SF transition (see \Figref{fig:SSE-Tdepend}). Moreover, we have shown with an analytical calculation of the AF SSE in the AF phase, that the sign reversal requires the following two conditions: (i) there exists a sizable N\'{e}el coupling $J_n$ at the AFI/M interface that produces the N\'{e}el spin current given by \Equref{eq:I_n}, and (ii) the spin dephasing $\tau_{\rm M}^{-1}$ of the M layer is not too strong and satisfies the inequality given by \Equref{eq:reversal-condition}. Our result indicates that the sign reversal of the AF SSE across the SF transition is not a generic property in a simple uniaxial antiferromagnet. We also argued that the existence of a sizable N\'{e}el coupling at the interface, which may manifest itself as a tendency to exhibit the exchange bias effect, is the key parameter for observing the sign reversal. These findings have identified important key parameters for the sign reversal problem in the AF SSE. We hope the present results contribute to further development of the AF spintronics.

%
%
\begin{acknowledgments}
This work was financially supported by JSPS KAKENHI Grant No. 19K05253, and by the Asahi Glass Foundation.
\end{acknowledgments}

\appendix
%
%

\section{\label{sec:phase-diagram} Calculation of the phase diagram}
In this Appendix, we analytically examine the phase diagram of the AFI modeled by \Equref{eq:Free_Energy1}.

\subsection{AF/SF boundary}

The phase diagram of the AFI is obtained by minimizing the free energy given in \Equref{eq:Free_Energy1}. By introducing the relative angle $\theta$ (angle $\phi$) between $\bm{m}$ and $\bm{n}$ ($\bm{n}$ and ${\bf \hat{x}}$) to represent $\bmn$ as $\bm{n}=n(\sin\theta \cos\phi, \, \sin\theta \sin\phi, \, \cos \theta)$, the free energy is rewritten in the dimensionless form as
\begin{eqnarray}
  \label{eq:Free_Energy_angle}
  F_{\mathrm{AFI}} &=& \frac{u_2}{2} n^2 + \frac{u_4}{4} n^4 + \frac{1}{2}(D+D') m^2 n^2 \nonumber\\
  && + \frac{1}{2}(K_0+K_1 \cos^2\phi-D m^2) n^2 \sin^2 \theta  \nonumber\\
  && + \frac{r_0}{2} m^2 - {H}_0 {m},
\end{eqnarray}
where we set $\bmm= m {\bf \hat{z}}$. Minimization of the free energy with respect to $m$, $n$, $\cos \phi$, and $\sin \theta$ gives their equilibrium values. Because of the in-plane anisotropy $K_1$, we see that $\cos \phi= 0$ (i.e., $\phi= \pi/2$) minimizes the free energy with respect to $\phi$.

From the term proportional to $\sin^2 \theta$ in the above free energy under the condition $\cos \phi =0$, we find that the sign of $K_0 - Dm^2$ determines the boundary between the AF phase ($\theta=0$) and the SF phase ($\theta= \pi/2$). To determine the critical field, we minimize the free energy with respect to $m$ and $n$ by calculating $\partial F_\text{AFI}/\partial m=0$ and $\partial F_\text{AFI}/\partial {n}=0$, and we obtain
\begin{gather}
\label{eq:m_minimization} {m} = \left\{ r_0+ (D\cos^2\theta+D') {n}^2 \right\}^{-1} {H}_0, \\
\label{eq:n_minimization} u_2 + u_4 {n}^2
+ (K_0-D {m}^2)\sin^2\theta + (D+D') {m}^2 = 0.
\end{gather}

We first determine the boundary by approaching from the AF phase ($\theta= 0$), where we have a condition $K_0 - D m^2 = 0_+$ at the boundary. By solving $D {m}^2=K_0$ with respect to $n$, we obtain 
\begin{equation}
  n^2 = \frac{1}{D+D'}\left(\sqrt{\frac{D}{K_0}} H_0-r_0\right). \label{eq:neq_AF}
\end{equation}
At the same time, substituting $m^2 = K_0/D$ into \Equref{eq:n_minimization}, we have
\begin{equation}
  u_2 + u_4 n^2 + \frac{D+D'}{D}K_0 = 0. \label{eq:mean_field_AF}
\end{equation}
By substituting \Equref{eq:neq_AF} into \Equref{eq:mean_field_AF} and solving for $H_0$, the critical field is calculated to be
\begin{equation}
    \label{eq:H_SF_upper}
  H_{\mathrm{SF}}^{(1)}
  = \sqrt{\frac{K_0}{D}}
  \left[
  r_0 -\frac{D+D'}{u_4}\left(u_2+\frac{D+D'}{D}K_0\right)
  \right].
\end{equation}

Next, we apply the same procedure to the SF phase ($\theta= \pi/2$), where we have a condition $K_0 - D m^2 = 0_-$ at the boundary. $D {m}^2=K_0$ with respect to $n$, we obtain
\begin{equation}
  n^2 = \frac{1}{D'} \left(\sqrt{\frac{D}{K_0}} H_0 -r_0 \right). \label{eq:neq_SF}
\end{equation}
At the same time, substituting $m^2=K_0/D$ into (\ref{eq:n_minimization}), we have
\begin{equation}
  u_2 + K_0 + u_4 n^2 + \frac{D'}{D}K_0 = 0. \label{eq:mean_field_SF}
\end{equation}
Substituting \Equref{eq:neq_SF} into \Equref{eq:mean_field_SF}, and solving for $H_0$, the critical field from the SF phase is given by
\begin{equation}
  \label{eq:H_SF_lower}
  H_{\mathrm{SF}}^{(2)}
  = \sqrt{\frac{K_0}{D}}
  \left[
  r_0 -\frac{D'}{u_4}\left(u_2+\frac{D+D'}{D}K_0\right)
  \right].
\end{equation}
In our numerical study, the position of the AF/SF boundary is well described by the average of $H_\text{SF}^{(1)}$ and $H_\text{SF}^{(2)}$, i.e., \Equref{eq:H_SF} in the main text. We have numerically confirmed that \Equref{eq:H_SF} reproduces the boundary obtained by free-energy minimization. 

\subsection{AF/PM and SF/PM boundaries}

We turn to the boundary between the AF phase and the PM phase. In the AF phase ($\theta=0$), we first substitute \Equref{eq:m_minimization} into \Equref{eq:Free_Energy_angle} to eliminate $m$, and then perform a Landau expansion with respect to $n$, obtaining 
\begin{flalign}
    F_{\mathrm{AFI}} &= \frac{1}{2r_0} \left[ u_2 r_0 + \frac{D+D'}{r_0} H_0^2 \right] n^2 \nonumber \\
  &+ \frac{1}{4r_0} \left[ u_4 r_0 -  2 \left(\frac{D+D'}{r_0}\right)^2 H_0^2 \right] n^4 + \cdots.
\end{flalign}
Since the coefficient of the quartic term is positive in the region in question, the AF phase and the PM phase are separated by a second-order transition. The boundary is determined by the coefficient of the quadratic term, i.e., $u_2 r_0^2 + (D+D'){H}_0^2 =0$, which gives the critical field given by \Equref{eq:H_AFPM} in the main text.

In a similar manner, we can also calculate the SF/PM phase boundary. In the SF phase ($\theta=\pi/2$),
we first substitute \Equref{eq:m_minimization} into \Equref{eq:Free_Energy_angle} to eliminate $m$, and then perform a Landau expansion with respect to $n$, obtaining 
\begin{eqnarray}
    F_{\mathrm{AFI}} &=& \frac{1}{2r_0} \left[ (u_2+K_0) r_0 + \frac{D'}{r_0} H_0^2 \right] n^2 \nonumber \\
  &+& \frac{1}{4r_0} \left[ u_4 r_0 -  2 \left(\frac{D'}{r_0}\right)^2 H_0^2 \right] n^4 + \cdots.
\end{eqnarray}
Since the coefficient of the quartic term is positive in the region in question, the SF phase and the PM phase are separated by a second-order transition. The boundary is determined by the coefficient of the quadratic term, i.e., $(u_2+K_0) r_0^2 + D'{H}_0^2 =0$, which gives the critical field given by \Equref{eq:H_SFPM} in the main text.

\section{\label{sec:AFMRmode} Calculation of the antiferromagnetic resonance modes}

In this Appendix, we analytically examine the AFMR modes in the AFI modeled by \Equref{eq:Free_Energy1}.

\subsection{AF phase}

Here, we focus on the AF phase ($\theta=0$) and calculate two AF magnon modes. In the AF phase, we can safely disregard the in-plane anisotropy $K_1$ as its effect on the magnons is negligibly small. We introduce $\delta \bm{m}=\bm{m} - \bm{m}_\text{eq}$ and $\delta \bm{n}=\bm{n} - \bm{n}_\text{eq}$, where $\bmm_{\rm eq}$ and $\bmn_{\rm eq}$ are the equilibrium values of $\bmm$ and $\bmn$. We first substitute these decompositions into Eqs. (\ref{eq:TDGL_m}) and (\ref{eq:TDGL_n}), and then linearize the equations with respect to $\delta \bmm$ and $\delta \bmn$. Introducing Fourier transformation $\delta O(t) = \int _{-\infty}^{\infty} \frac{d\omega}{2\pi} \delta O(\omega) e^{-i\omega t}$, we obtain the following eigenvalue equation:
\begin{align}
\label{eq:TDGL_matrix_AF}
(\omega-\hat{\mathcal{A}}_\text{AF})
\begin{pmatrix}
\delta m ^{-}(\omega) \\
\delta n ^{-}(\omega) \\
\end{pmatrix}
=
\begin{pmatrix}
0 \\
0 \\
\end{pmatrix},
\end{align}
where we have introduced rotating coordinate representation $\delta O^{\pm} = \delta O_x \pm i \delta O_y$ for a vector ${\bm O}$. The matrix $\hat{\mathcal{A}}_\text{AF}$ in the above equation is given by
\begin{equation}
  \hat{\mathcal{A}}_\text{AF}
  =
  \begin{pmatrix}
  a_\text{AF}, & b_\text{AF} \\
  c_\text{AF}, & d_\text{AF} \\
  \end{pmatrix},
\end{equation}
where each matrix component is defined as
\begin{equation}
  \begin{split}
    a_\text{AF}& = \widetilde{\omega}_0 H_0, \\
    b_\text{AF}&= \widetilde{\omega}_0 n_\text{eq} K_0, \\
    c_\text{AF}&= \widetilde{\omega}_0  n_\text{eq}(\widetilde{\chi}_\perp ^{-1} + D m_\text{eq}^2), \\
    d_\text{AF}&= \widetilde{\omega}_0 H_0 \widetilde{\chi}_\parallel \left( K_0 - D(m_\text{eq}^2-n_\text{eq}^2) \right),
  \end{split}
\end{equation}
where the imaginary part of the coefficient describing the relaxation is neglected, and $\widetilde{\chi}_\perp$ and $\widetilde{\chi}_\parallel$ are defined below \Equref{eq:AFMR}.

The equilibrium magnetization and staggered magnetization, $m_\text{eq}$ and $n_\text{eq}$, are calculated from Eqs.~(\ref{eq:m_minimization}) and (\ref{eq:n_minimization}) by setting $\theta=0$. From the condition that the determinant of \Equref{eq:TDGL_matrix_AF} is equal to zero, we can obtain the two magnon frequencies $\omega_\pm$ given in \Equref{eq:AFMR}.

\subsection{SF phase}
Here we focus on the SF phase ($\theta=\pi/2$) and calculate the eigenmodes. Following the same procedure as above, we can write a $6\times6$ matrix equation for $(\delta \bm{m},\delta \bm{n})^T $, but one notice that this matrix equation reduces to the following two matrix equations composed of $3\times3$ matrix:
\begin{align}
\label{eq:TDGL_matrix_QFMR}
(\omega-\hat{\mathcal{A}}_\text{QFMR})
\begin{pmatrix}
\delta m_x(\omega) \\
\delta m_y(\omega) \\
\delta n_z(\omega) \\
\end{pmatrix}
=
\begin{pmatrix}
0 \\
0 \\
0 \\
\end{pmatrix},
\end{align}
and
\begin{align}
\label{eq:TDGL_matrix_flat}
(\omega-\hat{\mathcal{A}}_\text{flat})
\begin{pmatrix}
\delta m_z(\omega) \\
\delta n_x(\omega) \\
\delta n_y(\omega) \\
\end{pmatrix}
=
\begin{pmatrix}
0 \\
0 \\
0 \\
\end{pmatrix},
\end{align}
where the matrix $\hat{\mathcal{A}}_\text{QFMR}$ and $\hat{\mathcal{A}}_\text{flat}$ are respectively given by
\begin{equation}
  \hat{\mathcal{A}}_\text{QFMR}
  =
  \begin{pmatrix}
  0          , & a_\text{SF}, & b_\text{SF} \\
  c_\text{SF}, & 0          ,           & 0 \\
  d_\text{SF}, & 0          ,           & 0 \\
  \end{pmatrix},
\end{equation}
and
\begin{equation}
  \hat{\mathcal{A}}_\text{flat}
  =
  \begin{pmatrix}
  0          , & e_\text{SF}, & 0           \\
  f_\text{SF}, & 0          , & g_\text{SF} \\
  0          , & h_\text{SF}, & 0           \\
  \end{pmatrix}.
\end{equation}
The matrix components of these matrices are defined by
\begin{equation}
  \begin{split}
    a_\text{SF}&= -i \widetilde{\omega}_0 H_0, \\
    b_\text{SF}&= -i \widetilde{\omega}_0  n_\text{eq} K_0, \\
    c_\text{SF}&=  i \widetilde{\omega}_0 H_0, \\
    d_\text{SF}&= -i \widetilde{\omega}_0 n_\text{eq} \widetilde{\chi}_\text{SF}^{-1}, \\
    e_\text{SF}&= -i \widetilde{\omega}_0 n_\text{eq} K_1, \\
    f_\text{SF}&=  i \widetilde{\omega}_0 n_\text{eq}(\widetilde{\chi}_\text{SF}^{-1}-2D'm_\text{eq}^2), \\
    g_\text{SF}&= 2i \widetilde{\omega}_0 n_\text{eq}^2 m_\text{eq}(D'-u_4), \\
    h_\text{SF}&=  i \widetilde{\omega}_0 m_\text{eq} K_1,
  \end{split}
\end{equation}
where we again neglect the relaxation term for simplicity. The equilibrium magnetization and staggered magnetization, $m_\text{eq}$ and $n_\text{eq}$, are calculated from Eqs. (\ref{eq:m_minimization}) and (\ref{eq:n_minimization}) by setting $\theta=\pi/2$, where $\widetilde{\chi}_\text{SF}=\left( r_0 + D'n_\text{eq}^2 \right)^{-1}$ is the spin susceptibility in the SF phase. The conditions for the determinant of \Equref{eq:TDGL_matrix_QFMR} and \Equref{eq:TDGL_matrix_flat} to become zero, respectively, give $\omega_{\rm QFMR}$ [\Equref{eq:QFMR}] and $\omega_{\rm flat}$ [\Equref{eq:flat-mode}].

\section{\label{sec:analytical_spin_current} Derivation of Eqs.~(\ref{eq:I_m}) and (\ref{eq:I_n})}
In this Appendix, we present the derivation of Eqs. (\ref{eq:I_m}) and (\ref{eq:I_n}). As stated in the main text, here we use dimensionful form of the TDGL equations (\ref{eq:TDGL_m}), (\ref{eq:TDGL_n}), and (\ref{eq:TDGL_s}). Since we are concerned with a low magnetic field region $H_0 \ll H_{\rm SF}$ in this calculation, we set $D=0$ in Eq.~(\ref{eq:TDGL_m}).

%
%
\subsection{\label{sec:analytical-I_m} Derivation of Eq.~(\ref{eq:I_m})}

We begin with the derivation of Eq. (\ref{eq:I_m}). This equation had already been calculated in the limit of $\omega_{\pm} \tau_\text{M} \ll 1$~\cite{Yamamoto2019}, and the calculation here is basically the same as Ref.~\cite{Yamamoto2019}. The additional aspect here is that we take into account the higher-order terms with respect to $\omega_\pm \tau_{\rm M}$.

The spin current injected into the M layer is defined by Eq.~(\ref{eq:Is}). By using the $z$ component of Bloch equation (\ref{eq:TDGL_s}) as well as introducing the rotating coordinate representation of the magnetization fluctuation $\delta m^\pm$ [see \Equref{eq:TDGL_matrix_AF} above], the spin current mediated by the magnetic coupling [\Equref{eq:couplingA}] is given by
\begin{align}
 \label{eq:analytical-I_m}
 I_s
 &= \frac{J_{m}}{\hbar} \int _{-\infty} ^{\infty} \frac{d\omega}{2\pi}
 \textrm{Im} \daverage{\delta m^- _{\omega} \delta s ^+ _{-\omega}},
\end{align}
where the Fourier transformation introduced above \Equref{eq:TDGL_matrix_AF} is used, and $\daverage{\delta m^-_{\omega} \delta s^+_{-\omega}}$ is defined by $\average{\delta m^- _{\omega} \delta s^+ _{\omega'}}=2\pi \delta(\omega+\omega') \daverage{\delta m^- _{\omega} \delta s^+ _{-\omega}}$.

The fluctuations $\delta m^-_{\omega}$ and $\delta s ^+ _{-\omega}$ can be obtained after linearizing the TDGL equations~(\ref{eq:TDGL_m}), (\ref{eq:TDGL_n}), and (\ref{eq:TDGL_s}) with respect to $\delta m^\pm_{\omega}$, $\delta n^\pm_{\omega}$, and $\delta s^\pm _{\omega}$:
\begin{align}
(\omega-\hat{\mathcal{A}}^\pm)
\begin{pmatrix}
\delta m _ \omega ^{\pm} \\
\delta n _ \omega ^{\pm} \\
\end{pmatrix}
&=
\begin{pmatrix}
i \xi ^{\pm} ({\omega}) \\
i \eta ^{\pm} ({\omega}) \\
\end{pmatrix} \nonumber \\
& \quad +
\begin{pmatrix}
\pm \frac{J_m}{\hbar}m_{\text{eq}}+i\frac{\Gamma_m J_m}{\gamma \hbar \mathfrak{h}_0} \\
\pm \frac{J_m }{\hbar}n_{\text{eq}} \\
\end{pmatrix} \delta s_{\omega}^{\pm},  \label{eq:TDGL_AF_Mcoupling} \\
\left( \omega + \frac{i}{\tau_{\rm M}} \right) \delta s^\pm_\omega
&=
i \frac{J_{m} \chi_{\rm M}}{\tau_{\rm M}} \delta m^\pm_\omega + i \zeta^\pm _\omega,
\label{eq:TDGL_M_Mcoupling}
\end{align}
where the above equations are also expanded up to the linear order with respect to $J_m$. The matrix $\hat{\mathcal{A}}^\pm$ is given by
\begin{equation}
\label{eq:matrix_component}
\hat{\mathcal{A}}^\pm
=
\begin{pmatrix}
a_\pm, & b_\pm \\
c_\pm, & d_\pm \\
\end{pmatrix},
\end{equation}
where the matrix component is defined by
\begin{align}
  a_\pm &=\pm \gamma H_0-i\Gamma _m \widetilde{\chi} _{\parallel} ^{-1},\\
  b_\pm &=\pm \gamma \mathfrak{h}_0 K_0 n_{\text{eq}}, \\
  c_\pm &=\pm \gamma \mathfrak{h}_0 \widetilde{\chi} _{\parallel} ^{-1} n_{\text{eq}},\\
  d_\pm &=\pm \gamma \mathfrak{h}_0 K_0 m_{\text{eq}}-i\Gamma _n K_0.
\end{align}

By operating the propagator $\hat{\mathcal{G}}^\pm=(\omega-\hat{\mathcal{A}}^\pm)^{-1}$ to \Equref{eq:TDGL_AF_Mcoupling} from the left-hand side, we obtain the dynamic fluctuations of $\delta m^\pm$, $\delta n^\pm$, and $\delta s^\pm$. Substituting $\delta m^\pm$, $\delta n^\pm$, and $\delta s^\pm$ thus obtained into \Equref{eq:analytical-I_m}, we obtain
\begin{equation}
  \label{eq:I_m-final}
  I_s = \frac{4k_B \Delta T {\chi}_{\rm M} J_m^2}{\epsilon_0 v_0 \hbar \tau_{\rm M}} \mathcal{L}_{m},
\end{equation}
where we used the expression for the correlator of thermal noises $\daverage{\xi_\omega ^- \xi_{-\omega}^+}=4k_\text{B} T \Gamma_m/(\epsilon_0 v_0)$, $\daverage{\eta_\omega ^- \eta_{-\omega}^+}=4k_\text{B} T \Gamma_n/(\epsilon_0 v_0)$, and $\daverage{\zeta_\omega ^- \zeta_{-\omega}^+}=4k_\text{B}(T+\Delta T) {\chi}_\text{M}\tau_\text{M}^{-1}$. In \Equref{eq:I_m-final}, $\mathcal{L}_{m}$ is defined by
\begin{equation}
  \label{eq:integral-L_m}
  \mathcal{L}_{m} =\int _{-\infty}^{\infty} \frac{d \omega}{2\pi}\Big[\Gamma _m |G_m^-(\omega)|^2 + \Gamma_n |G_n^-(\omega)|^2 \Big] \omega |g(\omega)|^2,
\end{equation}
where
\begin{equation}
  G_m^-(\omega) = \frac{\omega -d_-}{(\omega-\omega_{+}) (\omega-\omega_{-})},
\end{equation}
\begin{equation}
  G_n^-(\omega) = \frac{b_-}{(\omega-\omega_{+}) (\omega-\omega_{-})},
\end{equation}
and $g(\omega) = (\omega + i/\tau_{M})^{-1}$. 

We evaluate the frequency integral in \Equref{eq:integral-L_m} with residue theorem. By picking up two magnon poles at $\omega_+^*$ and $\omega_-^*$ in the upper-half plane with the asterisk denoting the complex conjugate, \Equref{eq:integral-L_m} becomes
\begin{eqnarray}
\label{eq:L_m}
\mathcal{L}_{m}
&=&
\frac{\mathcal{N}_{m1} + \tau_\text{M}^{-2}\mathcal{N}_{m2}}{\mathcal{D}\sqrt{Z^*}({\omega_+^*}^2+\tau_\text{M}^{-2})({\omega_-^*}^2+\tau_\text{M}^{-2})},
\end{eqnarray}
where $\mathcal{N}_{m1}$, $\mathcal{N}_{m2}$ and $\mathcal{D}$ are defined by
\begin{eqnarray}
  \mathcal{N}_{m1}
  &=&
  \{ \Gamma_m(\omega_+^*-d)(\omega_+^*-d^*)+\Gamma_nb^2 \} \nonumber\\
  &&\times \omega_+^*(XY+\Gamma_+\sqrt{Z^*}){\omega_-^*}^2  \nonumber\\
  &&+\{ \Gamma_m(\omega_-^*-d_-)(\omega_-^*-d_-^*)+\Gamma_n b_-^2 \} \nonumber\\
  &&\times \omega_-^* (-XY+\Gamma_+\sqrt{Z^*}){\omega_+^*}^2,   \label{eq:Nm1} \\
  \mathcal{N}_{m2}
  &=&
  \{ \Gamma_m(\omega_+^*-d_-)(\omega_+^*-d_-^*)+\Gamma_nb_-^2 \} \nonumber\\
  &&\times \omega_+^*(XY+\Gamma_+\sqrt{Z^*}) \nonumber\\
  &&+\{ \Gamma_m(\omega_-^*-d_-)(\omega_-^*-d_-^*)+\Gamma_n b_-^2 \} \nonumber\\
  &&\times \omega_- ^* (-XY+\Gamma_+\sqrt{Z^*}),  \label{eq:Nm2} \\
  \mathcal{D}&=&(\Gamma_+^2-Y^2)(X^2+\Gamma_+^2),  \label{eq:D}
\end{eqnarray}
and we introduced the notation $\Gamma_{\pm}=\Gamma_n K_0 \pm \Gamma_m \widetilde{\chi}_{\parallel}^{-1}$ and $\omega_+-\omega_- =\sqrt{Z} =X+iY$. Using the relation $X^2-Y^2 \approx (a_--d_-)^2+4b_-c_-=(\gamma H_0 )^2(1-\widetilde{\chi} _{\parallel} K_0)^2+4(\gamma h_0)^2 n_{\text{eq}}^2 \widetilde{\chi} _{\parallel} ^{-1} K_0$, $XY \approx (a_--d_-) \Gamma_-=\gamma H_0 (1-\widetilde{\chi} _{\parallel}  K_0)\Gamma _-$, $\omega_++\omega_-=a_-+d_-$, and $\omega_+ \omega_-=a_-d_--b_-c_-$, Eqs.(\ref{eq:Nm1})--(\ref{eq:D}) are calculated to be 
\begin{eqnarray}
  \label{eq:Nm1-final}
  \mathcal{N}_{m1}
  &=&
  \frac{1}{2}\omega_+^*\omega_-^*\sqrt{Z^*} \gamma \mathfrak{h}_0  m_{\text{eq}}K_0
  \widetilde{\chi}_{\parallel} \mathcal{D}, \\
  \mathcal{N}_{m2}
  &=&
  \label{eq:Nm2-final}
  \frac{1}{2}\sqrt{Z^*}\gamma \mathfrak{h}_0  m_{\text{eq}} \mathcal{D},
\end{eqnarray}
where we used
\begin{eqnarray}
  \label{eq:D-concrete}
  \mathcal{D}
  &=&4\Bigl\{\Gamma _m \widetilde{\chi} _{\parallel} ^{-1}\Gamma _n K_0(\gamma H_0)^2
  (1-\widetilde{\chi} _{\parallel}  K_0)^2 \nonumber\\
  &&+\Gamma _+ ^2(\gamma \mathfrak{h}_0 n_{\text{eq}})^2 \widetilde{\chi} _{\parallel} ^{-1} K_0 \Bigr\}
\end{eqnarray}
to simplify the result. Substituting Eqs.(\ref{eq:Nm1-final}) and (\ref{eq:Nm2-final}) into \Equref{eq:L_m}, the integral $\mathcal{L}_{m}$ now becomes
\begin{equation}
\mathcal{L}_{m}
= \frac{1}{2} \gamma \mathfrak{h}_0 m_{\text{eq}}
\frac{\omega_+ \omega_- K_0 \widetilde{\chi}_{\parallel}  + \tau_\text{M}^{-2}}{({\omega_+ }^2+\tau_\text{M}^{-2})({\omega_- }^2+\tau_\text{M}^{-2})} \label{eq:L_m-final}
\end{equation}
where we have used $\omega_+^* \approx \omega_+$ and $\omega_-^*\approx \omega_-$ since  $\text{Re}[\omega_+]\gg \text{Im}[\omega_+]$ and $\text{Re}[\omega_-]\gg \text{Im}[\omega_-]$. From \Equref{eq:I_m-final} and \Equref{eq:L_m-final} with the use of a relation $\epsilon_0 v_0 = \gamma \hbar \mathfrak{h}_0$, we finally obtain \Equref{eq:I_m}.

In passing, we shall discuss the formal relationship between the present approach and the other approach relying on the spin-mixing conductance phenomenology~\cite{Reitz2020}. According to Ref.~\cite{Reitz2020}, the spin pumping current is given by Eq.~(\ref{eq:Ispump_m}), where $g_m$ is the corresponding spin-mixing conductance. By comparing the spectral representation of $I_s$ in our approach [Eq.~(\ref{eq:I_m-final})] with Eq.~(\ref{eq:Ispump_m}), we identify $g_m= J_m^2 \chi_{\rm M} \tau_{\rm M}/\hbar$. Note that this correspondence holds only in the $|\omega_\pm| \tau_{\rm M} \to 0$ limit.

%
%
\subsection{\label{sec:analytical-I_n} Derivation of Eq.~(\ref{eq:I_n})}
We next present the derivation of \Equref{eq:I_n}. In the case of the N\'{e}el coupling, the spin current injected into the M layer is given by
\begin{align}
 \label{eq:analytical-I_n}
 I_s
  &= \frac{J_{n}}{\hbar} \int _{-\infty} ^{\infty} \frac{d\omega}{2\pi} \textrm{Im} \daverage{\delta n^- _{\omega} \delta s ^+ _{-\omega}},
\end{align}
where the spin current is mediated by the N\'{e}el coupling [\Equref{eq:couplingB}].
The fluctuations $\delta m^-_{\omega}$ and $\delta s ^+ _{-\omega}$ can be obtained in a manner similar to obtaining Eqs.~(\ref{eq:TDGL_AF_Mcoupling}) and (\ref{eq:TDGL_M_Mcoupling}), by taking notice of the replacement of $J_m$ with $J_n$ and resultant changes:
\begin{align}
(\omega-\hat{\mathcal{A}}^\pm)
\begin{pmatrix}
\delta m _ \omega ^{\pm} \\
\delta n _ \omega ^{\pm} \\
\end{pmatrix}
&=
\begin{pmatrix}
i \xi ^{\pm} ({\omega}) \\
i \eta ^{\pm} ({\omega}) \\
\end{pmatrix} \nonumber \\
& \quad +
\begin{pmatrix}
\pm \frac{J_n}{\hbar}n_{\text{eq}} \\
\pm \frac{J_n}{\hbar}m_{\text{eq}}+i\frac{\Gamma_n J_n}{\gamma \hbar \mathfrak{h}_0} \\
\end{pmatrix} \delta s_{\omega}^{\pm},  \label{eq:TDGL_AF_Ncoupling} \\
\left( \omega + \frac{i}{\tau_{\rm M}} \right) \delta s^\pm_\omega
&=
i \frac{J_n \chi_{\rm M}}{\tau_{\rm M}} \delta n^\pm_\omega + i \zeta^\pm _\omega.
\label{eq:TDGL_M_Ncoupling}
\end{align}

By substituting $\delta m^\pm$, $\delta n^\pm$, and $\delta s^\pm$ obtained from Eqs.~(\ref{eq:TDGL_AF_Ncoupling}) and (\ref{eq:TDGL_M_Ncoupling}) into \Equref{eq:analytical-I_n}, we have
\begin{equation}
  \label{eq:I_n-final}
  I_s = \frac{4k_B \Delta T {\chi}_{\rm M} J_n^2}{\epsilon_0 v_0 \hbar \tau_{\rm M}} \mathcal{L}_{n},
\end{equation}
where $\mathcal{L}_{n}$ is defined by
\begin{equation}
  \label{eq:integral-L_n}
  \mathcal{L}_{n} =\int _{-\infty}^{\infty} \frac{d \omega}{2\pi}\Big[\Gamma _m |F_n^-(\omega)|^2 + \Gamma_n |F_m^-(\omega)|^2 \Big] \omega  |g(\omega)|^2,
\end{equation}
in which we defined
\begin{equation}
  F_n^-(\omega) = \frac{c_-}{(\omega-\omega_{+}) (\omega-\omega_{-})},
\end{equation}
and
\begin{equation}
  F_m^-(\omega) = \frac{\omega -a_-}{(\omega-\omega_{+}) (\omega-\omega_{-})}.
\end{equation}

We evaluate the frequency integral in \Equref{eq:integral-L_n} with residue theorem. By picking up two magnon poles $\omega_+^*$ and $\omega_-^*$ in the upper-half plane, we have
\begin{eqnarray}
\label{eq:L_n}
\mathcal{L}_{n}
&=&
\frac{
\Gamma_mc_-^2 + \Gamma_n (\omega_+^*-a_-)(\omega_+^*-a_-^*)
}
     {
(\Gamma_+-Y)(X+i\Gamma_+)\sqrt{Z^*}({\omega_+^*}^2+\tau_\text{M}^{-2})
}\omega_+^* \nonumber\\
&&+
\frac{
\Gamma_mc_-^2 + \Gamma_n(\omega_-^*-a_-)(\omega_-^*-a_-^*)
}
{
(\Gamma_++Y)(X-i\Gamma_+)\sqrt{Z^*}({\omega_-^*}^2+\tau_\text{M}^{-2})
}\omega_-^* \nonumber\\
&=&
\frac{\mathcal{N}_{{n1}} + \tau_\text{M}^{-2}\mathcal{N}_{{n2}}}{\mathcal{D}\sqrt{Z^*}({\omega_+^*}^2+\tau_\text{M}^{-2})({\omega_-^*}^2+\tau_\text{M}^{-2})},
\end{eqnarray}
where
\begin{eqnarray}
  \mathcal{N}_{{n1}}
  &=&
  \{ \Gamma_mc_-^2 + \Gamma_n(\omega_+^*-a_-)(\omega_+^*-a_-^*) \} \nonumber\\
  &&\times \omega_+^*(XY+\Gamma_+\sqrt{Z^*}){\omega_-^*}^2  \nonumber\\
  &&+\{ \Gamma_m c_-^2 + \Gamma_n(\omega_-^*-a_-)(\omega_-^*-a_-^*) \} \nonumber\\
  &&\times \omega_-^* (-XY+\Gamma_+\sqrt{Z^*}){\omega_+^*}^2,   \label{eq:Nn1}\\
  \mathcal{N}_{{n2}}
  &=&
  \{ \Gamma_m c_-^2 + \Gamma_n(\omega_+^*-a_-)(\omega_+^*-a_-^*) \} \nonumber\\
  &&\times \omega_+^*(XY+\Gamma_+\sqrt{Z^*}) \nonumber\\
  &&+\{ \Gamma_m c_-^2 + \Gamma_n (\omega_-^*-a_-)(\omega_-^*-a_-^*)\} \nonumber\\
  &&\times \omega_-^* (-XY+\Gamma_+\sqrt{Z^*}),   \label{eq:Nn2}
\end{eqnarray}
and $\mathcal{D}$ is defined in \Equref{eq:D}. Again using the relation $X^2-Y^2 \approx (a_--d_-)^2+4b_-c_-=(\gamma H_0 )^2(1-\widetilde{\chi} _{\parallel} K_0)^2+4(\gamma h_0)^2 n_{\text{eq}}^2 \widetilde{\chi}_{\parallel}^{-1} K_0$, $XY \approx (a_--d_-)\Gamma_-=\gamma H_0 (1-\widetilde{\chi} _{\parallel}  K_0)\Gamma _-$, $\omega_++\omega_- = a_- +d_-$, and $\omega_+ \omega_- =a_-d_--b_-c_-$, Eqs.(\ref{eq:Nm1}) and (\ref{eq:Nm2}) become
\begin{eqnarray}
  \mathcal{N}_{{n1}}
  &=&
  \frac{1}{2}\omega_+^*\omega_-^*\sqrt{Z^*} \gamma \mathfrak{h}_0  m_{\text{eq}}(K_0\widetilde{\chi}_{\parallel})^{-1} \mathcal{D}, \label{eq:Nn1-final} \\
  \mathcal{N}_{{n2}}
  &=&
    \frac{1}{2}\sqrt{Z^*}\gamma \mathfrak{h}_0  m_{\text{eq}} \mathcal{D}. \label{eq:Nn2-final}
\end{eqnarray}
Note that $\mathcal{N}_{{n2}}$ in \Equref{eq:Nn2-final} becomes equal to $\mathcal{N}_{m2}$ in \Equref{eq:Nm2-final}. Substituting Eqs.(\ref{eq:Nn1-final}) and (\ref{eq:Nn2-final}) into \Equref{eq:L_n}, we now have
\begin{equation}
\label{eq:L_n-final}
\mathcal{L}_{n} = \frac{1}{2} \gamma \mathfrak{h}_0 m_{\text{eq}} \frac{\omega_+ \omega_- (K_0 \widetilde{\chi}_{\parallel})^{-1}  + \tau_\text{M}^{-2}}{({\omega_+ }^2+\tau_\text{M}^{-2})({\omega_- }^2+\tau_\text{M}^{-2})},\\
\end{equation}
where we have again used $\omega_+^* \approx \omega_+$ and $\omega_-^*\approx \omega_-$. Using \Equref{eq:I_n-final}, (\ref{eq:L_n-final}), and a relation $\epsilon_0 v_0 = \gamma \hbar \mathfrak{h}_0$, we finally obtain \Equref{eq:I_n}.

As in the previous section, we discuss the formal relationship between the present approach and the other approach relying on the spin-mixing conductance phenomenology~\cite{Reitz2020}. According to Ref.~\cite{Reitz2020}, the spin pumping current is given by Eq.~(\ref{eq:Ispump_n}), where $g_n$ is the corresponding spin-mixing conductance. By comparing the spectral representation of $I_s$ in our approach [Eq.~(\ref{eq:I_n-final})] with Eq.~(\ref{eq:Ispump_n}), we identify $g_n= J_n^2 \chi_{\rm M} \tau_{\rm M}/\hbar$. Note that this correspondence holds only in the $|\omega_\pm| \tau_{\rm M} \to 0$ limit.

%
%

%
%


%

\end{document}